\documentclass[prd,superscriptaddress,onecolumn,nofootinbib,aps,tightenlines,preprintnumbers,
12pt]{revtex4}

\usepackage{epsfig}
\usepackage{amsmath}
\usepackage{amsfonts}
\usepackage{dsfont}
\usepackage{fouridx}
\usepackage{hyperref}
\usepackage{wasysym}
\usepackage{epstopdf}
\usepackage{subfig}
\usepackage[justification=raggedright]{caption}

\begin{document}

\preprint{\hbox{CALT-TH-2014-144}  }

\title{Supermassive Black Holes from Ultra-Strongly Self-Interacting Dark Matter}

\author{Jason Pollack}
\email{jpollack@caltech.edu}
\affiliation{Walter Burke Institute for Theoretical Physics, Division of Physics, Mathematics \& Astronomy, California Institute of Technology, Pasadena, CA 91125}
\author{David N. Spergel}
\affiliation{Department of Astrophysical Sciences, Peyton Hall, Princeton University, Princeton, NJ 08544}
\author{Paul J. Steinhardt}
\affiliation{Department of Physics and Princeton Center for Theoretical Science, Princeton University, Princeton, NJ 08544}

\date{\today}

\begin{abstract}
%original abstract
%If all dark matter is elastically self-interacting, there exist stringent constraints on its self-interaction cross section per unit mass, $\sigma\apprle1\ \mathrm{cm^{2}/g}$. 
%We point out that if only some fraction $f\ll1$ of the dark matter is self-interacting, most observational constraints are rendered irrelevant, even if this subdominant component is ultra-strongly self-interacting, with $\sigma\gg1\ \mathrm{cm^{2}/g}$.
%This ultra-strongly self-interacting dark matter (uSIDM) component can form seed black holes in the center of halos if it undergoes gravothermal collapse.
%We find that uSIDM halos form black holes with mass a few percent of the total uSIDM mass in several hundred halo interaction times. 
%For reasonable values of $\sigma f$, black holes can form at high enough redshifts to grow to $\sim10^{9}M_{\odot}$ quasars by $z\apprge6$ by accreting baryons, alleviating tension within the standard $\Lambda\mathrm{CDM}$ cosmology.
%The ubiquitous formation of central black holes in halos could create cores in dwarf galaxies by blowing out matter during binary black hole mergers, potentially resolving the ``too big to fail" problem.
We consider the cosmological consequences if a small fraction ($f\apprle 0.1$) of the dark matter is ultra-strongly self-interacting, with an elastic self-interaction cross-section per unit mass $\sigma\gg1\ \mathrm{cm^{2}/g}$.   
This possibility evades all current constraints that assume that the self-interacting component makes up the majority of the dark matter.    
Nevertheless, even a small fraction of ultra-strongly self-interacting dark matter (uSIDM) can have observable consequences on astrophysical scales.  
In particular, the uSIDM subcomponent can undergo gravothermal collapse and form seed black holes in the center of a halo.
These seed black holes, which form within several hundred halo interaction times, contain a few percent of the total uSIDM mass in the halo.
For reasonable values of $\sigma f$, these black holes can form at high enough redshifts to grow to $\sim10^9 M_\odot$ quasars by $z \apprge 6$, alleviating tension within the standard $\Lambda$CDM cosmology. 
The ubiquitous formation of central black holes in halos could also create cores in dwarf galaxies by ejecting matter during binary black hole mergers, potentially resolving the ``too big to fail" problem.
\end{abstract}

\widetext
\maketitle
\vfill\eject

\tableofcontents
%\vfill\eject

\baselineskip=14pt

%%%%%%%%%%%%%%%%%%%%%%%%%%%%%%%%%%%%
\section{Introduction}

Although $\mathrm{\Lambda CDM}$ cosmology provides an excellent fit to the observational data on $\apprge\mathrm{Mpc}$ scales \cite{Ade:2013zuv}, its success is less certain over the strongly nonlinear, $\apprle\mathrm{kpc}$ regime relevant to the substructure within galactic halos. 
The deviation of galactic cores from an expected cuspy density profile \cite{Moore:1994yx,Moore:1997sg} and an apparent shortfall of observed Milky Way satellites relative to expectations from simulations \cite{Klypin:1999uc,Moore:1999nt} originally motivated considerations that the dark matter might have non-negligible self-interactions \cite{Spergel:1999mh}. 
Although a combination of improved theoretical understanding and additional observations had appeared to alleviate these problems and remove the phenomenologically interesting parameter space for self-interacting dark matter (SIDM) \cite{Yoshida:2000uw,Markevitch:2003at,Gnedin:2000ea}, a recent reevaluation of the constraints \cite{Rocha:2012jg,Peter:2012jh} has demonstrated that SIDM with an velocity-independent elastic self-interaction cross section per unit mass $\sigma\simeq0.1-1\ \mathrm{cm^{2}/g}\simeq0.2-2\ \mathrm{b/GeV}$ can simultaneously meet all constraints and alleviate the discrepancies between $\mathrm{\Lambda CDM}$ and observations.

In this paper, we exhibit a distinct area of the SIDM parameter space which is likewise both allowed by observations and potentially interesting phenomenologically. 
In particular, we examine the case in which most of the dark matter remains non-self-interacting (or weakly self-interacting) as in the standard $\mathrm{\Lambda CDM}$ picture, but a small fraction $f\ll1$ of the dark matter is made up of a subdominant component that is \textit{ultra-strongly self-interacting}, abbreviated as uSIDM, with $\sigma\gg1\ \mathrm{cm^{2}/g}$ (where $\sigma$ denotes the cross section \emph{per unit mass}). 
Because most of the dark matter remains inert, constraints that rely on distinguishing the overall behavior of SIDM halos from their $\mathrm{CDM}$ counterparts are no longer relevant. 

Consider, for example, the constraints placed on the SIDM cross section from observations of the Bullet Cluster (1E 0657-6). 
Observations reveal an offset between the gas ``bullet" and the dark matter centroid of the currently merging subcluster.
Under the assumption that the subcluster has already passed through the main cluster, this offset is due to stripping and deceleration of gas in the subcluster due to interactions with the main cluster itself.
The observation that the dark matter has not been slowed to the same degree allows limits to be placed on the dark matter self-interaction cross section.
The strongest constraint \cite{Randall:2007ph} comes from the measurement of the ratio of mass-to-light ratios of the subcluster and the main cluster, which is found to be $0.84\pm0.07$. 
Under the assumption that the subcluster and main cluster had the same initial mass-to-light ratio before merger, this means that the subcluster cannot have lost more than 23\% of its mass.

In \cite{Randall:2007ph}, this measurement plus estimates of the subcluster escape velocity and merger speed were used to constrain $\sigma\apprle0.6\ \mathrm{cm^2/g}$ when $f=1$.
However, it is clear that, even in the extreme example that $all$ of the SIDM mass in the subcluster was lost to scattering, current observations would not be able to detect the SIDM subcomponent if $f<0.07$, within the uncertainty on the mass-to-light ratio. 
So constraints from the Bullet Cluster certainly do not apply when $f\ll10^{-1}$, regardless of the size of the self-interaction cross section per unit mass $\sigma$. 
Even when $f\sim0.1$, $\sigma$ may not be well-constrained, since one pass through the main cluster would not suffice to strip all of the SIDM from the bullet.

We note that observations of another cluster undergoing a major merger, A520 \cite{Mahdavi:2007yp,Jee:2012sr,Clowe:2012am,Jee:2014hja} have not provided similar constraints on the SIDM cross section; here the dark matter centroid of the subcluster is in fact coincident with the (presumably stripped) gas.
Under certain assumptions, this can be taken as evidence of a nonzero dark matter self-interaction cross section per unit mass, as strong as $0.94\pm0.06\ \mathrm{cm^2/g}$ in the latest observations \cite{Jee:2014hja}. 
The limited number of ongoing major merger events in the observable universe makes it hard to give an overall estimate of the self-interaction cross section from major mergers, but future surveys could potentially combine many minor merger events to measure $\sigma$ with a precision of $0.1\ \mathrm{cm^2/g}$
\cite{Harvey:2013tfa}.

Regardless of the situation for $f=1$ SIDM, we have seen that there are no observational constraints on a uSDIM component of the dark matter with $f\apprle 0.1$. 
At the same time, of course, a small component of uSIDM by itself is unable to produce cores or dissolve substructure to any observable degree. 
We point out, though, that a uSIDM component%
%\footnote{We already know that the dark matter is multi-component: neutrinos make up $10^{-2.9\pm0.1}$ of the critical energy density of the universe \cite{Fukugita:2004ee}, or around $1/250$ of the dark matter.
%In this paper we will often set $f=0.01$, so the energy density in the SIDM will be of the same order of magnitude.%
%}%
\ of the dark matter could instead explain another potential discrepancy with the $\mathrm{\Lambda CDM}$ picture: the existence of billion-solar-mass quasars at high redshifts $z\apprge6.5-7$ (for reviews, see \cite{Dokuchaev:2007mf, Volonteri:2010wz, Sesana:2011qi, Treister:2011yi, Kelly:2011ab, 2013ASSL..396..293H}). 
In \textsection2 we review the observational situation and the difficulties with explaining it within $\mathrm{\Lambda CDM}$. 
In \textsection3 we suggest an alternative: gravothermal collapse of an ultra-strongly self-interacting dark matter component. We review the mechanism of gravothermal collapse, specialize to the case of a halo containing uSIDM, and solve the problem numerically. 
We apply the results of \textsection3 to individual observations of high-redshift quasars in \textsection4, then discuss broader cosmological implications in \textsection5, including a potential way for uSIDM to indirectly produce cores in dwarf halos. 
We finally conclude in \textsection6.

%%%%%%%%%%%%%%%%%%%%%%%%%%%%%%%%%%%%
\section{Supermassive Black Holes}\label{sec:smbh}

Supermassive black holes (SMBHs) which grow primarily via gas accretion are Eddington-limited: the gravitational force on the accreting gas is balanced by its own radiation pressure. 
Hence growth via gas accretion cannot proceed faster than exponentially, with an $e$-folding rate bounded by the inverse of the Salpeter time \cite{Salpeter:1964kb}:
\begin{equation}
\label{eq:tsal}
t_{\mathrm{Sal}}=\frac{\epsilon_{r}\sigma_{T}c}{4\pi Gm_{p}}\approx\left(\frac{\epsilon_{r}}{0.1}\right)45.1\ \mathrm{Myr},
\end{equation}
where $\sigma_{T}$ is the Thompson cross section,
\begin{equation}
\sigma_{T}=\frac{8\pi}{3}\left(\frac{e^{2}}{4\pi\epsilon_{0}m_{e}c^{2}}\right)^{2},
\end{equation}
$m_p$ and $m_e$ are respectively the proton and electron masses, and $\epsilon_r$ is the radiative efficiency, which ranges from $1-\sqrt{8/9}\approx0.057$ to $1-\sqrt{1/3}\approx0.42$ as the angular momentum of the black hole increases from zero to its extremal value \cite{Shapiro:2004ud}; in astrophysical applications, $\epsilon_r$ is typically taken to be $\epsilon_r=0.1$.
Accretion faster than the Eddington limit, $\dot{M_\mathrm{Edd}}=M t_\mathrm{Sal}^{-1}$, onto a black hole of mass $M$ will result in a radiation pressure exceeding the gravitational force, driving outflows which should quickly halt this excessive accretion.
Yet several dozen quasars with masses a few $\times\ 10^{9}\ M_{\odot}$ have been detected at redshifts $z\apprge6$, including a quasar, ULAS J1120+0641, with mass $2.0_{-0.7}^{+1.5}\times10^{9}\ M_{\odot}$ at redshift $z=7.085$ \cite{Mortlock:2011va,Venemans:2012dt}. 
Using the Planck Collaboration's best-fit cosmological values \cite{Ade:2013zuv}, $z=7.085$ corresponds to $747\ \mathrm{Myr}$ after the Big Bang, so even continuous Eddington accretion since the Big Bang can only increase the mass of a seed black hole by a factor of $1.6\times10^{7}$. 
If we make the standard assumption that black hole seeds are formed from Pop III stars, the seed cannot have formed before around $z\sim30$, so the maximum growth factor shrinks by another order of magnitude, to $1.75\times10^{6}$, requiring%
%\footnote{The WMAP9 cosmology \cite{Hinshaw:2012aka} improves this result, but only slightly: the %growth factor is $2.37\times10^{6}$.%
%}%
\ a seed black hole mass $\sim10^{3}M_{\odot}$. 

More generally, in order to explain the observed abundance of $\sim1/\mathrm{Gpc^{3}}$ billion-solar-mass quasars at $z\simeq6$ \cite{2013ASSL..396..293H} within $\Lambda\mathrm{CDM}$, we must form $10^{2-3}M_{\odot}$ seed black holes soon after the beginning of baryonic structure formation and grow these black holes continuously at near-Eddington rates for $\sim800\ \mathrm{Myr}$.
Some simulations have shown this can be achieved \cite{Li:2006ti}, but only by making optimistic assumptions about cooling and star formation \cite{Tegmark:1996yt,Gao:2006ug}, fragmentation \cite{Turk:2009ae,Stacy:2009zt,McKee:2007yx}, photoevacuation \cite{Johnson:2006gd,Abel:2006gw,Yoshida:2006dv}, black hole spin \cite{Bardeen:1972fi,Zhang:1997dy,Narayan:2011eb}, and black hole mergers \cite{1983MNRAS.203.1049F,Merritt:2004xa,Haiman:2004ve}. 
We emphasize, in particular, that these results depend critically on the assumption of $\epsilon_r=0.1$; because the $e$-folding time itself depends linearly on the radiative efficiency, the maximum mass formed by a given time is \textit{exponentially} sensitive to its value.
Because quasar masses are inferred by measuring their luminosities and assuming they are Eddington-limited, increasing the assumed radiative efficiency will decrease the inferred quasar mass by $\epsilon_r^{-1}$. 
However, this reduction in required mass is made negligible by the much larger number of $e$-folds required to reach it. 
Recent work, both theoretical \cite{Shapiro:2004ud} and observational \cite{Trakhtenbrot:2014dza}, has found $\epsilon_r\gtrsim0.2$, which would be catastrophically incompatible with an assumption of black hole growth driven by Eddington accretion. 

One alternative is to allow for extended periods of super-Eddington gas accretion. 
Super-Eddington accretion is known to be possible, for example when outflows of gas and radiation are collimated \cite{Shakura:1972te, Jiang:2014tpa}, and extended periods of super-Eddington growth could account for the observed supermassive high-redshift quasars \cite{Volonteri:2014lja, Madau:2014pta}.
However, estimates of quasar masses and luminosities at low redshifts using emission line widths indicate that, at least in the late universe, the vast majority of quasars are constrained to radiate at the Eddington limit \cite{Kollmeier:2005cw}, or possibly well below it \cite{Steinhardt:2009ig,Steinhardt:2011wr}.

In this paper, we will therefore neglect the possibility of extended super-Eddington accretion.
We will assume that growth of black holes from baryonic accretion is limited to exponential growth with an $e$-folding time given by the Salpeter time (\ref{eq:tsal}).
In order to facilitate comparision of uSIDM to the standard picture, we will, however, allow for continuous accretion of baryons at this limit once a seed black hole has formed, despite the potential issues mentioned in the previous paragraph.
In other words, we attempt to modify the mechanism by which black hole seeds are formed, while leaving the simplest conventional mechanism for their growth from seeds to supermassive black holes intact.
It would be easy to combine our results with more realistic baryon accretion histories.

Finally, we note that future observations in the near-infrared, e.g.\ with the James Webb Space Telescope and Wide Field Infrared Survey Telescope (WFIRST), and in the radio, e.g.\ with the Square Kilometer Array, should be able to detect (or place limits on the density of) even intermediate-mass ($\sim10^5M_\odot$) quasars out to $z\sim10$ \cite{Haiman:1997bv, Haiman:2000ky, Haiman:2004ny,Whalen:2012ib}, providing vastly more information about the formation and growth of high-redshift quasars.

%%%%%%%%%%%%%%%%%%%%%%%%%%%%%%%%%%%%
\section{Gravothermal Collapse}\label{sec:gravothermal_collapse}

Motivated by the tensions within the standard ($\Lambda\mathrm{CDM}$) picture discussed in the previous section, we propose an alternative mechanism for black hole seed formation: the gravothermal collapse \cite{LyndenBell:1968yw} of the uSIDM component of a dark matter galactic halo. 
The simplest form of gravothermal collapse occurs in a population of gravitating point particles with elastic short-range interactions.
The classic illustration of the mechanism is globular clusters, where the point particles are stars. 
Stellar short-range interactions are not purely elastic, so in this case collapse is eventually halted by binary formation.
A gas of SIDM, however, has only elastic interactions, so core collapse continues until relativistic instability results in the formation of a black hole, which promptly Bondi accretes \cite{Bondi:1952ni} the optically thick core of SIDM that surrounds it. 

In this section we make this intuitive picture precise.
Full expressions will be given below, but in brief we find that the uSIDM component of a galactic halo undergoes gravothermal collapse in $\sim460$ halo relaxation times, forming a black hole which contains $\sim2\%$ of the uSIDM mass of the galaxy. 
The halo relaxation time is a complicated expression which depends on the halo mass and time of formation as well as the uSIDM properties, but we show in the following sections that, for reasonable values of uSIDM fraction $f$ and cross section per unit mass $\sigma$, there exist halos that can easily form seed black holes, and grow them using uSIDM and baryons, to achieve $10^{9}M_{\odot}$ SMBHs by redshift 6. 

Before formulating the problem, we first review the gravothermal collapse mechanism itself.
Intuitively, gravothermal collapse depends on the simple observation that gravitationally bound systems have negative specific heat. 
For a virialized system, this is immediate:
\begin{equation}
0=2T+V=T+E\rightarrow E=-T.
\end{equation}
Now consider two systems, an inner, gravitationally bound system with negative specific heat and an outer system surrounding it with positive specific heat---the inner and outer parts of a globular cluster, for example. 
Evolution towards equilibrium will direct both mass and heat outward, causing both the inner and the outer system to increase in temperature. 
A possible physical mechanism is a two-body scattering in the inner system which sends one star closer to the core (where it gains potential energy and thus speeds up, increasing the temperature of the inner system) and kicks one star out to the periphery (where its higher speed increases the temperature of the outer system). 
Importantly, we see that the inner system \emph{shrinks} as it heats up. 

Now two outcomes are possible, depending on the specific heat of the two systems as a function of their masses. 
If the outer system always has the smaller (magnitude of) specific heat, its temperature will eventually grow to exceed that of the inner system, and the entire assemblage of masses will reach equilibrium. 
On the other hand, if the outer system grows in mass too quickly, its specific heat will become too large and its temperature will never catch up to the inner system. Hence the inner system will continue shrinking in mass and growing in temperature until the thermodynamic description breaks down. 
This is precisely the \emph{gravothermal catastrophe }\cite{LyndenBell:1968yw}. 
In the case of a globular cluster (at least an idealized one with uniform-mass stars), the gravothermal collapse process is halted by binary formation, which acts as an energy sink \cite{Heggie:1975tg,Hut:1992wz}. 
If the uSIDM interacts purely via elastic scattering, however, no bound state formation is possible, and gravothermal collapse can drive the core to relativistic velocities, where it undergoes catastrophic collapse into a black hole via the radial instability \cite{1966SvA.....9..742Z,1985ApJ...298...34S,1985ApJ...298...58S,1986ApJ...307..575S}.

%%%%%%%%%%%%%%%%%%%%%%%%%%%%%%%%%%%%
\subsection{The Gravothermal Fluid Equations}

We now consider the gravothermal collapse of a general two-component dark matter halo, where the self-interacting component comprises some fraction $f$ of the mass of the halo. 
At this stage we do not yet specialize to the uSIDM case, with $f\ll1$. 
To avoid confusion, we will therefore refer to the two different components of the halo as SIDM (making up a fraction $f$ of the total mass of the halo) and (ordinary) CDM (making up the remainder), denoting the SIDM as uSIDM only when $f\ll1$.
To simulate the collapse, we employ the gravothermal fluid approximation \cite{1980MNRAS.191..483L,Balberg:2002ue,2011MNRAS.415.1125K}, which reduces the problem to a set of coupled partial differential equations that can then be solved numerically. 
First consider the general case for an $f=1$ fluid, i.e.\ a halo composed entirely of SIDM. 
A spherically symmetric ideal gas of point particles in hydrostatic equilibrium with arbitrary conductivity $\kappa$ obeys the following equations \cite{1980MNRAS.191..483L}:
\begin{equation}
\frac{\partial M}{\partial r}=4\pi r^{2}\rho\label{eq:fund1}
\end{equation}
\begin{equation}
\frac{\partial\left(\rho\nu^{2}\right)}{\partial r}=-\frac{GM\rho}{r^{2}}\label{eq:fund2}
\end{equation}
\begin{equation}
\frac{L}{4\pi r^{2}}=-\kappa\frac{\partial T}{\partial r}\label{eq:fund3}
\end{equation}
\begin{equation}
\frac{\partial L}{\partial r}=-4\pi\rho r^{2}\nu^{2}\left(\frac{\partial}{\partial t}\right)_{M}\ln\frac{\nu^{3}}{\rho},\label{eq:fund4}
\end{equation}
where $\nu(r)$ is the one-dimensional velocity dispersion and $L(r)$ the total heat radiated \textit{inward} through a sphere of radius $r$. 
The first equation (\ref{eq:fund1}) simply defines the integrated mass distribution in terms of the density. 
The second (\ref{eq:fund2}) is the statement of hydrostatic equilibrium: we inserted Euler's equation into the Poisson equation for a spherically symmetric potential and used the equation of state for an ideal gas, $p=\rho\nu^{2}$. 
The third (\ref{eq:fund3}) states that the heat flux is proportional to the temperature gradient, with proportionality constant given by the conductivity $\kappa$. 
The fourth (\ref{eq:fund4}) is the second law of thermodynamics, inserting the specific entropy of an ideal gas of point particles $u=\frac{k_{\mathrm{B}}}{m}\ln(\frac{T^{3/2}}{\rho})$ and using the relation $\nu^{2}=k_{\mathrm{B}}T/m$. 
This gives a set of four differential equations with four dependent variables $\{M,\ \rho,\ \nu,\ L\}$ and two independent variables $\{r,\ t\}$.
(The temperature $T$ is directly related to $\nu$ by $\nu^{2}=k_{\mathrm{B}}T/m$.)

To make progress, we need an expression for the form of the thermal conductivity $\kappa$ in terms of our physical parameter, the elastic scattering cross section per unit mass $\sigma$.
Dimensional analysis alone will not suffice: we have one time scale, the fluid relaxation time 
\begin{equation}
t_{r}\equiv1/(a\rho\sigma\nu),
\end{equation}
 with $a=\sqrt{16/\pi}\approx2.257$ for hard-sphere interactions, but two length scales, the mean free path $\lambda\equiv1/(\rho\sigma)$ and the Jeans length or gravitational scale height $H\equiv\sqrt{\nu^{2}/(4\pi G\rho)}$.
Following \cite{Balberg:2002ue,2011MNRAS.415.1125K}, we find the unique length scales in the two limiting cases, the short mean free path (smfp) regime $\lambda\ll H\rightarrow\ell_\mathrm{smfp}=\lambda$ and the long mean free path (lmfp) regime $\lambda\gg H\rightarrow\ell_{lmfp}=H$, and combine them in reciprocal to get a final length scale, $\ell\equiv\left(\ell_\mathrm{smfp}^{-1}+\ell_\mathrm{lmfp}^{-1}\right)^{-1}$. 
In the smfp regime, transport theory tells us that 
\begin{equation}
\frac{L}{4\pi r^{2}}\approx-\frac{3}{2}a^{-1}b\rho\frac{\lambda^{2}}{t_{r}}\frac{\partial\nu^{2}}{\partial r}.
\end{equation}
The coefficient $b$ is calculated perturbatively in Chapman-Enskog theory%
%\footnote{Balberg, Shapiro, and Inagaki \cite{Balberg:2002ue} erroneously quote %\textbf{$b=25\pi/\left(32\sqrt{6}\right)\approx1.002$ }(first pointed out in Koda and Shapiro %\cite{2011MNRAS.415.1125K}, who give the correct result).%
%}%
\ \cite{1981phki.book.....L}, \textbf{$b=25\sqrt{\pi}/32\approx1.385$}.
In the lmfp regime, the flux equation is well approximated as
\begin{equation}
\frac{L}{4\pi r^{2}}\approx-\frac{3}{2}C\rho\frac{H^{2}}{t_{r}}\frac{\partial\nu^{2}}{\partial r},
\end{equation}
where $C$ is a constant setting the scale on which the two conduction mechanisms are equally effective, determined by N-body simulations \cite{2011MNRAS.415.1125K} to be $C\approx290/385\approx0.75$. 
Hence the final expression is
\begin{equation}
\frac{L}{4\pi r^{2}}=-\frac{3}{2}ab\nu\sigma\left[a\sigma^{2}+\frac{b}{C}\frac{4\pi G}{\rho\nu^{2}}\right]^{-1}\frac{\partial\nu^{2}}{\partial r}.
\end{equation}

Now consider the more general case, $f\ne1$. Hydrostatic equilibrium is separately satisfied for each species of particle%
%\footnote{This can be thought of as a manifestation of Dalton's law: the two species have no short-range %interactions, so the total pressure is equal to the sum of the partial pressures and equilibrium of the %entire ensemble requires both fluids to be individually in equilibrium.%
%}
, but the gravitational potential is of course sourced by both species, giving the coupling between the two components. 
Because the non-SIDM component is taken to be collisionless, it has $\sigma=0$, so $L^{ni}=0$.
So the total system is governed by six partial differential equations with six dependent variables $\{M,\ \rho^{int},\ \rho^{ni},\ \nu^{int},\ \nu^{ni},\ L^{int}\}$ and two independent variables $\{r,\ t\}$:
\begin{equation}
\frac{\partial M}{\partial r}=4\pi r^{2}\left(\rho^{int}+\rho^{ni}\right)\label{eq:master 1}
\end{equation}
\begin{equation}
\frac{\partial\left(\rho^{int}\left(\nu^{int}\right)^{2}\right)}{\partial r}=-\frac{GM\rho^{int}}{r^{2}}\label{eq:master he}
\end{equation}
\begin{equation}
\frac{\partial\left(\rho^{ni}\left(\nu^{ni}\right)^{2}\right)}{\partial r}=-\frac{GM\rho^{ni}}{r^{2}}
\end{equation}
\begin{equation}
\frac{L^{int}}{4\pi r^{2}}=-\frac{3}{2}ab\nu^{int}\sigma\left[a\sigma^{2}+\frac{b}{C}\frac{4\pi G}{\rho^{int}\left(\nu^{int}\right)^{2}}\right]^{-1}\frac{\partial\left(\nu^{int}\right)^{2}}{\partial r}\label{eq:master luminosity}
\end{equation}
\begin{equation}
\frac{\partial L^{int}}{\partial r}=-4\pi\rho^{int}r^{2}\left(\nu^{int}\right)^{2}\left(\frac{\partial}{\partial t}\right)_{M}\ln\frac{\left(\nu^{int}\right)^{3}}{\rho^{int}}
\end{equation}
\begin{equation}
0=\left(\frac{\partial}{\partial t}\right)_{M}\ln\frac{\left(\nu^{ni}\right)^{3}}{\rho^{ni}}.\label{eq:master 6}
\end{equation}
As before, the first equation gives the total mass distribution, while the second and third enforce hydrostatic equilibrium. 
The fourth determines how the SIDM fluid conducts heat and the fifth how the flux gradient affects the fluid. 
Finally, the sixth equation ensures that the entropy of the collisionless component is conserved, $3\dot{\nu}/\nu=\dot{\rho}/\rho$. 
Notice that the fraction $f$ does not appear in the differential equations themselves, but only in the boundary conditions: we must have 
\begin{equation}
\frac{\int_{0}^{\infty}4\pi r^{\prime2}\rho^{int}(r^{\prime})dr^{\prime}}{\int_{0}^{\infty}4\pi r^{\prime2}\rho^{ni}(r^{\prime})dr^{\prime}}=\frac{f}{1-f}
\end{equation}
at all times.

%%%%%%%%%%%%%%%%%%%%%%%%%%%%%%%%%%%%
\subsection{Initial Conditions}\label{sub:initial}

In principle (\ref{eq:master 1} --\ref{eq:master 6}) can be solved exactly given appropriate boundary conditions at $r=0$ and $r=\infty$ and a set of initial radial profiles which obey the equations. 
In practice, this is computationally impossible: even finding the initial profiles for an arbitrary $\sigma$ is infeasible. 
Balberg, Shapiro, and Inagaki \cite{Balberg:2002ue}, considering the $f=1$ case, took the $\sigma\rightarrow0$ limit, which admits a self-similar solution where separation of variables is possible, then found the eigenvalues of the resulting system of ordinary spatial differential equations and took the resulting profiles as their initial conditions for the more general $\sigma\ne0$ case. 

We will instead \textit{assume} that SIDM self-interactions are unimportant during the process of halo formation, so that the the SIDM and collisionless components have the same initial profile. 
This allows us to use the results of (collisionless) $\mathrm{\Lambda CDM}$ simulations. 
We simplify further by approximating the initial halo by an NFW profile, 
\begin{equation}
\rho_{\mathrm{NFW}}(r)=\frac{\rho_{s}}{(r/r_{s})(1+r/r_{s})^{2}},\label{eq:nfw profile}
\end{equation}
where $\rho_{s}$ and $r_{s}$ are the characteristic density and scale radius, respectively. 
Since the NFW profile has a characteristic radius, we can state our assumption more precisely: we assume that halo formation proceeds much faster than heat conduction, which is true when the dynamical timescale of collapse is much less than the relaxation timescale due to collisions:
\begin{equation}
t_{\mathrm{dyn}}(r_{s})\ll t_{\mathrm{rel}}(r_{s})\approx\frac{1}{\tau_{s}}t_{\mathrm{dyn}}(r_{s})\rightarrow\tau_{s}\ll1;
\end{equation}
i.e.\ so long as the halo is \textit{optically thin at its characteristic radius}.
Again, if the optical depth is small, 
\begin{equation}
\label{optically_thin}
\tau\equiv f\rho_{\mathrm{NFW}}r\sigma\ll1\rightarrow
\sigma f\le\frac{1}{\rho_{s}r_{s}},
\end{equation}
typical SIDM particles have not yet undergone any self-interaction by the time of halo formation, so we are justified in assuming they follow the same initial profile as the collisionless dark matter, $\rho_{0}^{int}(r)=f\rho_{\mathrm{NFW}}(r)$. 

Before checking the validity of this assumption, we comment on the consequences of taking a different initial profile. 
The NFW profile is particularly simple: its form means that the optical depth at small radii, $r\ll r_{s}$, is independent of radius, so a small characteristic optical depth implies that the central regions are also optically thin despite the presence of a cusp. 
Modern $\mathrm{\Lambda CDM}$ simulations, however, have tended to find density profiles more complicated than the NFW profile. 
Profiles with cores or at least less cuspy behavior, e.g.\ Einasto profiles \cite{Merritt:2005xc,Graham:2006ae}, will have $\tau\ll1$ everywhere if $\tau_{s}\ll1$. 
Below we will see that SIDM halos with initial NFW density profiles grow cores on a scale of tens of halo relaxation times anyway, so shallower initial profiles will only result in slightly smaller times before black hole formation. 
Profiles with more cuspy behavior, e.g.\ generalized NFW or Zhao profiles \cite{Zhao:1995cp} with inner slope $\alpha\apprge1$, will unavoidably have regions at very small radii in the optically thick regime. 
Below we will see that SIDM halos with initial NFW profiles first evacuate the cusp to form cores before beginning the gravothermal collapse process, and it seems reasonable to conclude that the same thing will happen for non-pathological cuspier profiles.
We conclude that imposing a different profile should not significantly change the behavior investigated below.

When is the assumption that $\tau_{s}\ll1$ justified? 
Recall that the characteristic radius $\rho_{s}$ and radius $r_{s}$ for an NFW profile are given in terms of the halo virial mass $M_{\Delta}$ and concentration $c$:
\begin{equation}
r_{\Delta}\equiv cr_{s},
\end{equation}
\begin{equation}
M_{\Delta}\equiv M(r_{\Delta})=\int_{0}^{r_{\Delta}}4\pi r^{2}\rho_{\mathrm{NFW}}(r)dr=4\pi\rho_{s}r_{s}^{3}\left[\ln(1+c)-\frac{c}{1+c}\right],
\end{equation}
\begin{equation}
\rho_{s}\equiv\delta_{c}\rho_{crit}(z).
\end{equation}
The density contrast $\delta_c$ is in turn given by 
\begin{equation}
\delta_{c}=\frac{\Delta}{3}\frac{c^{3}}{K_{c}},  
\end{equation}
where $K_{c}\equiv\ln(1+c)-c/(1+c)$.
The problem thus reduces to finding an expression for $\Delta$, the virial overdensity.
In the spherical collapse model, this is given by $\Delta\sim18\pi^2\Omega_{m}^{0.45}$ for a flat universe \cite{Lahav:1991wc,Eke:1996ds,Bryan:1997dn,Neto:2007vq}; $\Delta$ hence approaches the familiar value of 178 in the matter-dominated era.

Inserting these expressions into (\ref{optically_thin}) above yields an inequality for $\sigma f$ in terms of $c$ and $M_\Delta$, along with the redshift of virialization $z$:
%\begin{equation}
%\begin{array}{ccc}
%\sigma f\le\frac{1}{\rho_{s}r_{s}} & = & (4\pi)^{-1/3}M_{\Delta}^{-1/3}\left(\frac{\Delta\rho_{crit}(z)}{3}\right)^{-2/3}Kc^{-2}\\
% & = & 24.56\ \mathrm{cm^{2}/g}\times\left(\frac{M_{\Delta}}{10^{12}M_{\odot}}\right)^{-1/3}\times\left(\frac{\rho_{crit}(z)}{\rho_{crit}(z=15)}\right)^{-2/3}Kc^{2}
%\end{array}
%\end{equation}
\begin{equation}
\sigma f\le\frac{1}{\rho_{s}r_{s}}=(4\pi)^{-1/3}M_{\Delta}^{-1/3}\left(\frac{\Delta\rho_{crit}(z)}{3}\right)^{-2/3}K_c\,c^{-2}
\label{eq:opt_thin}
\end{equation}
 
\begin{equation}
\label{optically_thin_bound}
=24.56\ \mathrm{cm^{2}/g}\times\left(\frac{M_{\Delta}}{10^{12}M_{\odot}}\right)^{-1/3}\times\left(\frac{\rho_{crit}(z)}{\rho_{crit}(z=15)}\right)^{-2/3}K_c\,c^{-2}.
\end{equation}
In the second line we have inserted the typical halo parameters we will consider below: $z=15$, $M_\Delta=10^{12} M_\odot$.

It remains to insert plausible values for the concentration $c$.
Individual halos of mass $M_\Delta$ formed at a fixed redshift $z$ will have varying concentrations, but there should be some mass- and redshift-dependent median concentration, $c(M_\Delta,z)$.
Prada \textit{et al.} \cite{Prada:2011jf} used the Millennium \cite{Springel:2005nw,BoylanKolchin:2009nc}, Bolshoi \cite{Klypin:2010qw}, and MultiDark \cite{Riebe:2011gp} simulations to examine the shape of the $c(M_\Delta,z)$ curve with varying mass and redshifts.
They found that for each redshift considered (from $z\sim0-6$) the concentration formed a U-shaped curve: it was minimized at a certain value of the mass, but increased steeply both above and below this mass.
Furthermore, they found that both the minimum value of the concentration and the mass at which this minimum was realized decreased with increasing redshift.
At the large redshifts we consider, the cluster-sized halos needed to form supermassive black holes are far more massive than the bottom of the U-shaped curve; accordingly, the fitting formulae given in \cite{Prada:2011jf} predict that the concentration for these halos will be extremely large, of the order of $c\sim10^5$ for the halo parameters above.
If this were true, the initial density profiles of these large, early halos would be extremely concentrated, so that their inner regions are extremely thick even for $\sigma f\apprge10^{-6}$. 
In this case the simulations presented in this paper would not be reliable.

%Eke, Navarro, and Steinmetz \cite{Eke:2000av} performed simulations to study the dependence of $c$ on the halo mass $M_{\Delta}$ and the underlying cosmology: we use their publicly available fitting formulas%
%\footnote{\url{http://www.astro.uvic.ca/~jfn/cens/}.%
%} and the Planck cosmological parameters \cite{Ade:2013zuv} to evaluate $c(M_{\Delta})$ and thus determine the upper bound on $\sigma$ as a function of $M_{\Delta}$ and $f$. \textbf{{[}Quote for range of redshifts and masses. 
%Expect $\sigma f=$(tens to hundreds) but need to fix the code to properly account for nonzero redshifts.{]}
%}\ 

We emphasize, however, that the fitting formulae of \cite{Prada:2011jf} were devised using simulated halos only out to $z\sim6$; they should not be trusted so far away from their domain of validity.
Accordingly, we have consulted the high-redshift halo catalogs of the FIRE simulation \cite{Hopkins:2013vha}, which attempted to resolve an overdense region at high redshift.
The catalogs use the Amiga Halo Finder \cite{Knollmann:2009pb} to measure $c$ in the same way as defined in \cite{Prada:2011jf}. 
We are interested in the concentration parameters of the most massive halos formed at a given redshift.
Perhaps unsurprisingly, we find that, even at $z\sim30$, halo concentrations range from 2 to 11, similar to the values found at lower redshifts in the simulations consulted in \cite{Prada:2011jf}, rather than the much higher values predicted by naively applying the fitting formulae.
We do not attempt to construct the full $c(M_\Delta,z)$ curve at high redshifts on the basis of this limited data, but we do assume that realistic halos will take concentrations in this observed range.

The upper bound on $\sigma f$ for which $\tau_{s}\le1$ ranges from $0.32-2.65\ \mathrm{cm^{2}/g}$ as concentrations decreasing from 11 to 2 are inserted into (\ref{optically_thin_bound}).
In the remainder of this paper we will typically set $c=9$, which gives a bound of $0.425\ \mathrm{cm^{2}/g}$.
In Section \ref{discussion} below we will find that this bound is of the same order of magnitude as the cross section needed to produce the desired high-redshift supermassive black holes using uSIDM. 
Accordingly, there is a surprisingly small region of parameter space where both the assumption of an initial NFW profile is valid and the desired black holes are produced.
We will discuss this further in Section \ref{discussion}.
For now, we note only that the qualitative results of this paper should still hold even when our assumption of an initial NFW profile is invalid. 
Outside of this range, we expect that gravothermal collapse should still occur---in fact, it should occur \emph{faster} because core formation will have begun even before virialization---but the particular expressions given here will no longer be valid.

%%%%%%%%%%%%%%%%%%%%%%%%%%%%%%%%%%%%
\subsection{Integration of the Equations}\label{sub:integration}

\begin{figure}
  \begin{center}
    \includegraphics[width=1\textwidth]{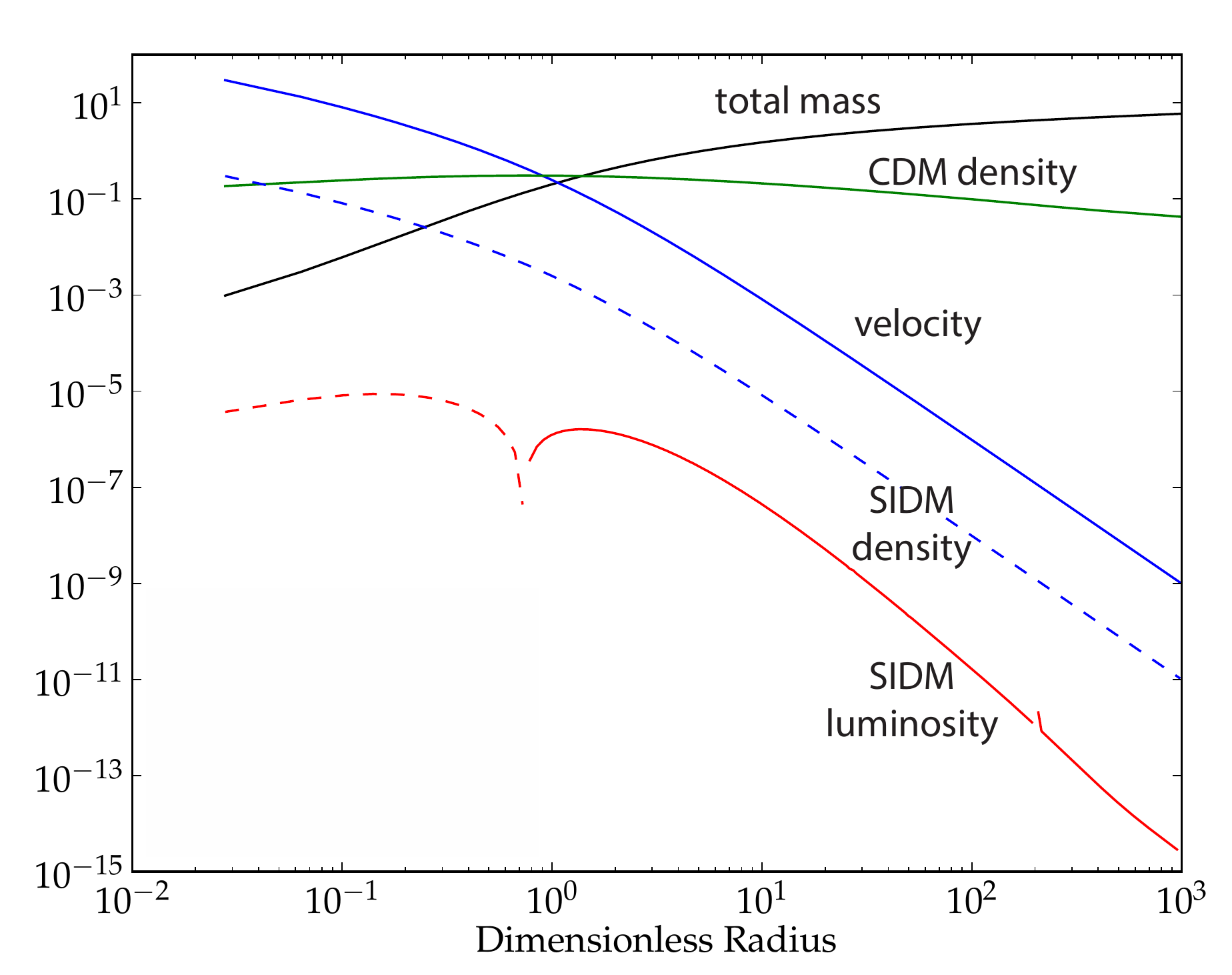}
  \end{center}
  \caption{The dimensionless initial profiles for an NFW halo, with $f=0.01$.
  The CDM and SIDM have the same velocity profile, and their density profiles have the same shape but a normalization differing by $f/(1+f)$.
  As expected for SIDM, the initial luminosity at small radii is negative (shown as dashed on the plot), indicating that the cusp is being forced outward as a core begins to form. 
  The glitch in the luminosity at $\tilde{r}=200$ is a numerical artifact. 
  \label{fig:nfw init}}
\end{figure}

Given the initial conditions, we can proceed to integrate the system of equations (\ref{eq:master 1} --\ref{eq:master 6}). 
We first move to a dimensionless form of the problem by choosing fiducial mass and length scales $\{M_{0},\ R_{0}\}$. 
Then the remaining dependent variables are given naturally in terms of these quantities, e.g.\ $\nu_{0}=\sqrt{GM_{0}/R_{0}}$.
Full expressions for all dependent variables in terms of $M_0$ and $R_0$ are given in Section 5 of \cite{Balberg:2002ue}.
The cross section per unit mass is now expressed dimensionlessly by $\hat{\sigma}=\sigma/\sigma_{0},\ \sigma_{0}=4\pi R_{0}^{2}/M_{0}$.
It is convenient to use the two quantities already specified in the NFW profile, $\{\rho_{s},\ r_{s}\}$; we therefore take 
\begin{equation}
\{M_{0}=4\pi R_{0}^{3}\rho_{s},\ R_{0}=r_{s}\}.
\end{equation}
Note that we have made a different choice of $\{M_{0},\ R_{0}\}$ than \cite{Balberg:2002ue}, since we consider a cuspy NFW profile rather than a cored one and thus work with characteristic rather than central quantities.
Finally, the timescale is set by the initial relaxation time at the characteristic radius, 
\begin{equation}
t_{r,c}(0)=1/(fa\rho_{s}\nu_{s}\sigma),
\end{equation}
so the independent variable can also be made dimensionless.
Dimensionless quantities are written with tildes (e.g.\ $\tilde{\rho}, \tilde{t}$).
The resulting initial profiles for an $f=0.01$ halo are shown in Figure \ref{fig:nfw init}. 

We solve the problem by spatially discretizing into $N$ concentric spherical shells, initially evenly logarithmically spaced in radius. 
At each timestep, we first apply the effects of heat conduction, which increases the energy within each shell, then adjust the profile to maintain hydrostatic equilibrium.
The heat conduction step is simple: we determine the luminosity profile from the density and velocity dispersion using the dimensionless, discretized form of (\ref{eq:master luminosity}), then adjust the energy of each shell accordingly (using $dU\equiv L dt$ for a finite but small timestep).
Timesteps are chosen so that the change of (dimensionless) specific energy $\tilde{u_{i}}=3\tilde{\nu}_{i}^{2}/2$ is not large: we require $\Delta\tilde{u}_{i}/\tilde{u}_{i}<\varepsilon\ll1$ for each shell $i$, typically taking $\varepsilon=0.001$. 
This means that as the gravothermal catastrophe approaches and core temperatures and densities become large, the size of timesteps will decrease dramatically: as expected, we cannot integrate through the collapse because the fluid approximation itself breaks down there. 

To carry out the hydrostatic equilibrium step, we use the method of Lagrangian zones, in which the radius of each shell is adjusted while the mass it contains is left constant. 
The relaxation process, which involves long-range gravitational interactions rather than heat conduction via collisions, is entropy-preserving%
%\footnote{We are able to make this assumption because the dynamical timescale of the system is much %less than the thermal timescale of conduction, so we can think of the system as continuously returning %to hydrostatic equilibrium.%
%}
, so it preserves the adiabatic invariants $A_i\equiv\tilde{\rho}_i \tilde{V}_i^{5/3}$ for each shell $i$. After the heat conduction step, each shell is temporarily out of hydrostatic equilibrium, so that the equality (\ref{eq:master he}) is violated by some amount $\Delta_i$. 
The problem is to adjust the density, velocity dispersion, and radius of each shell $i$, such that hydrostatic equilibrium is again satisfied ($\Delta_i=0\ \forall i$) while preserving the adiabatic invariants. 
The assumption of adiabaticity, along with the use of Lagrangian zones to keep the mass of each shell fixed, fixes the density and velocity changes as functions of the set of changes of radii $\Delta\tilde{r}_i$.
Hence the requirement of hydrostatic equilibrium gives a system of differential equations for the changes of radii which, when linearized, is tridiagonal (since the thickness of each shell depends not only on its own central radius but that of its nearest neighbors). 
The resulting system is solved using a standard linear algebra library%
\footnote{\url{http://www.gnu.org/software/gsl}.%
}. 

%%%%%%%%%%%%%%%%%%%%%%%%%%%%%%%%%%%%
\subsection{Results}

\begin{figure}
  \begin{center}
    \subfloat[\label{fig:nfw_frac_density.pdf}][]{\includegraphics[width=.7\textwidth]{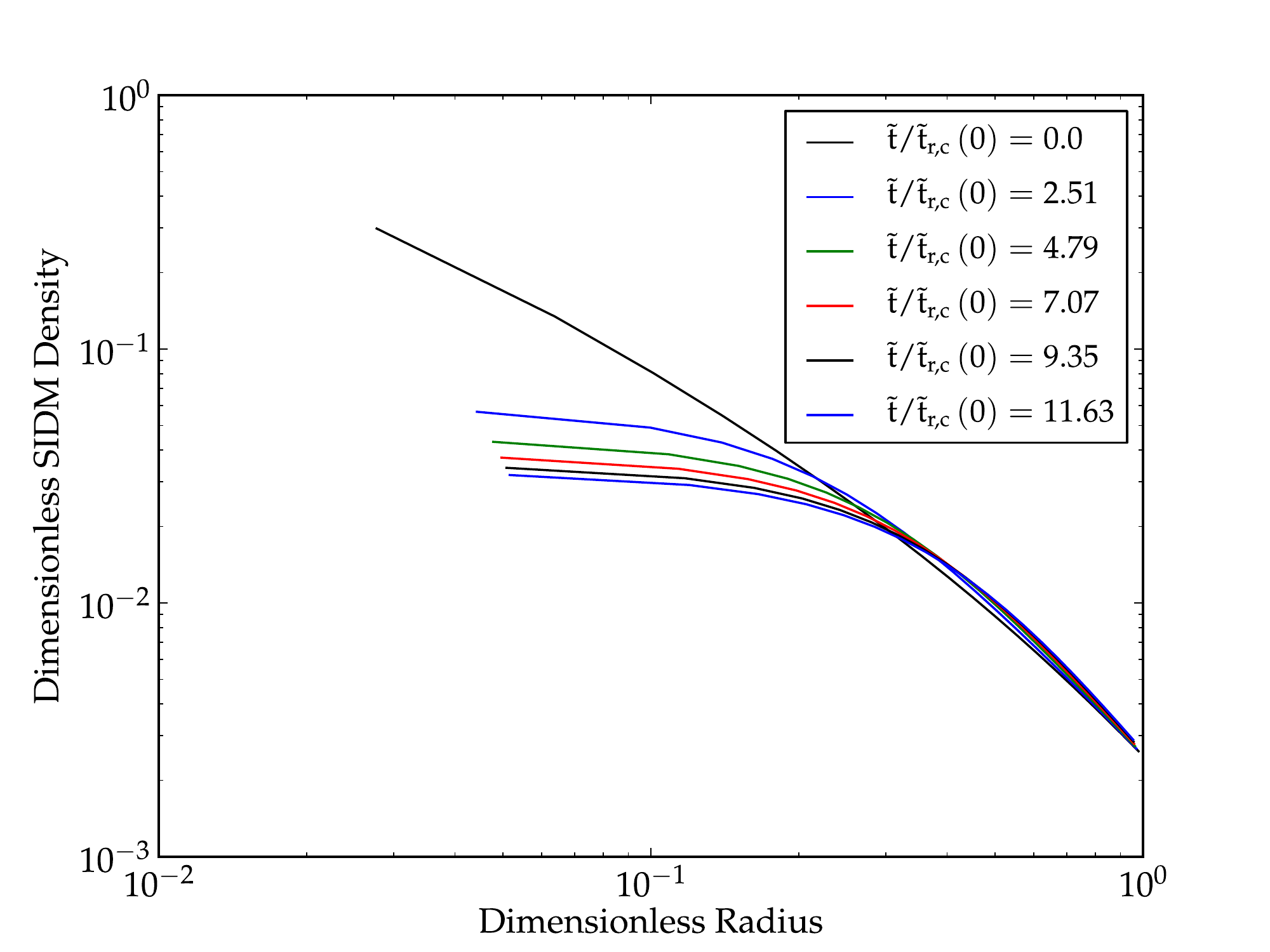}}
    \\
    \subfloat[\label{fig:nfw_density.pdf}][]{\includegraphics[width=.7\textwidth]{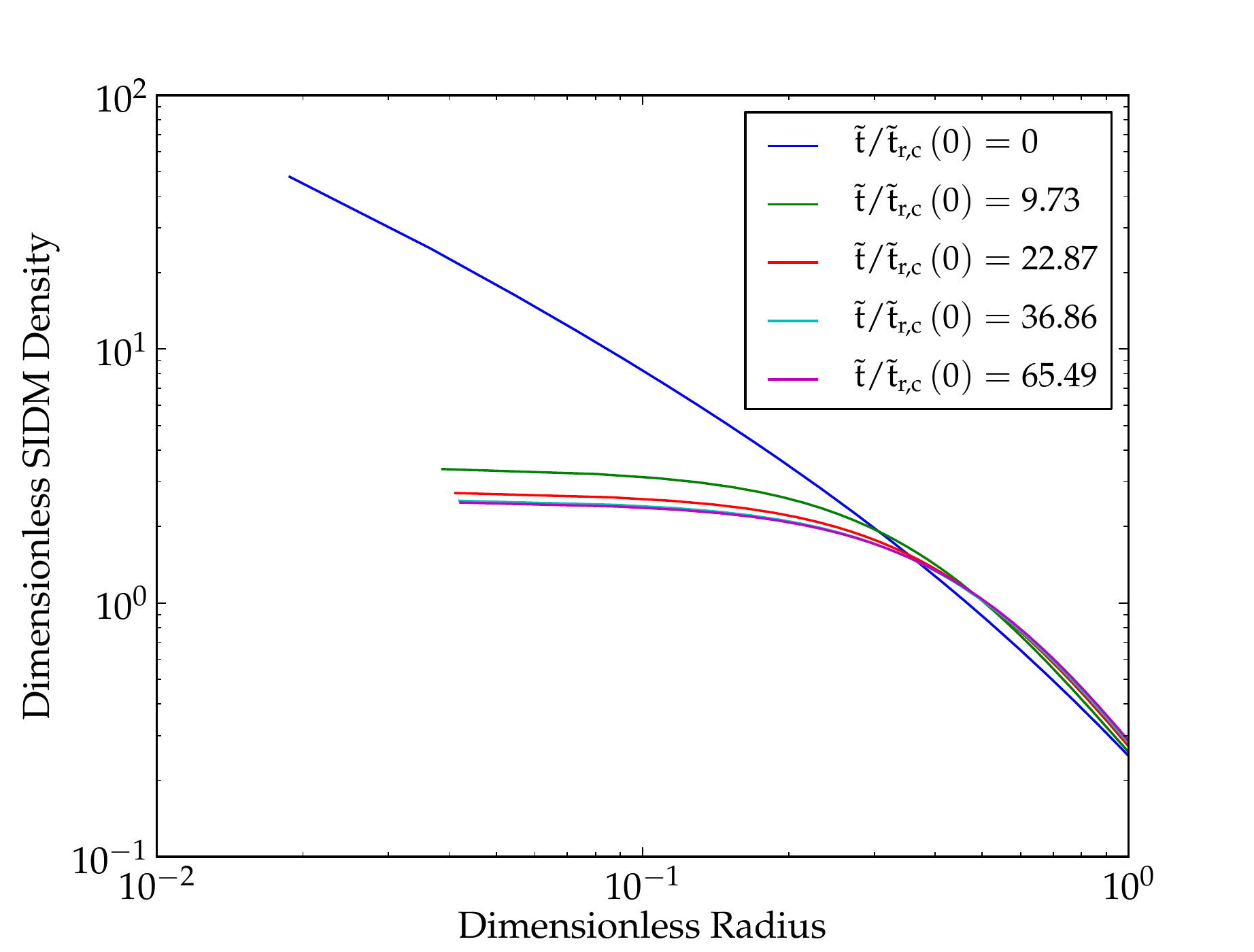}}
  \end{center}
  \caption{(a) Evolution of SIDM density profiles, starting with an $f=0.01$ NFW halo. 
Only the inner part of the halo is shown; the outer part still asymptotes to $r^{-3}$ as in Figure \ref{fig:nfw init} above for all halos. 
From top to bottom, profiles are at $0.0$, $2.51$, $4.79$, $7.07$, $9.35$, and $11.63$ central relaxation times.
Because $t_{r}\propto\rho^{-1}$, this corresponds to integrating for $\sim1000$ relaxation times in an $f=1$ halo. 
However, comparison to the $f=1$ results below suggests that the density profile flattens in the same manner, just $f^{-1}$ times slower: evidently the non-interacting dark matter has little influence on the central SIDM evolution. 
(b) Evolution of an $f=1$ halo starting from NFW initial conditions. 
For clarity, only the inner portion of the density profile is shown: the outer profile has not yet changed significantly at this stage. 
From top to bottom, profiles are at $0.0$, $9.73$, $22.87$, $36.86$, and $65.49$ central relaxation times.
As in the $f=0.01$ case, the density profile is flattening as a core develops. 
\label{fig:initial evolution}}
\end{figure}

Unfortunately, the above procedure is still insufficient to integrate (\ref{eq:master 1}--\ref{eq:master 6}) in full generality. 
The problem is that, because the SIDM and collisionless dark matter are \emph{separately} in hydrostatic equilibrium, the method of Lagrangian zones will result in different sets of radii for the two species. 
But in order to perform subsequent timesteps, we need the total mass distribution at each radius for both types of DM. 
For computationally feasible numbers of shells ($N\sim400$), interpolation is not accurate enough to preserve numerical stability and the distributions cannot be integrated all the way up to the point of gravothermal collapse. 

We can, however, consider the two limiting cases.
(Luckily, these happen to be the cases we are interested in!)
In the pure SIDM case $f=1$, there is only one species and the problem does not arise. 
In the uSIDM case, $f\ll1$, we can ignore the gravitational backreaction of the uSIDM component on the collisionless DM and assume that it maintains an NFW profile throughout, allowing the calculation of its mass distribution analytically at every point. 
We expect that the two cases should yield similar results, because the temporary violation of (\ref{eq:master he}), the hydrostatic equilibrium condition, after each heat conduction timestep is overwhelmingly due to the increase on the LHS of the equation, from heat conduction, rather than from interactions with the collisionless component, on the RHS of the equation. 
This is just the statement that the self-interaction is much larger than gravitational strength. 
We indeed find that this is the case, at least qualitatively. 
Consider Figure \ref{fig:initial evolution}, which shows the early evolution of $f=0.01$ and $f=1$ halos with the same value of $\hat{\sigma}$. 
We see that behavior is indeed qualitatively the same: in both cases, a core begins to form as heat conduction dissolves the initial cusp. 
Note that the time scales are different: in the uSIDM case the relaxation time is increased by a factor of $f^{-1}$ since the uSIDM density is a factor of $f$ lower. 
So Figure \ref{fig:initial evolution} suggests that uSIDM evolution is the same as the $f=1$ case, just $f^{-1}$ times slower. 

\begin{figure}
\includegraphics[width=1\textwidth]{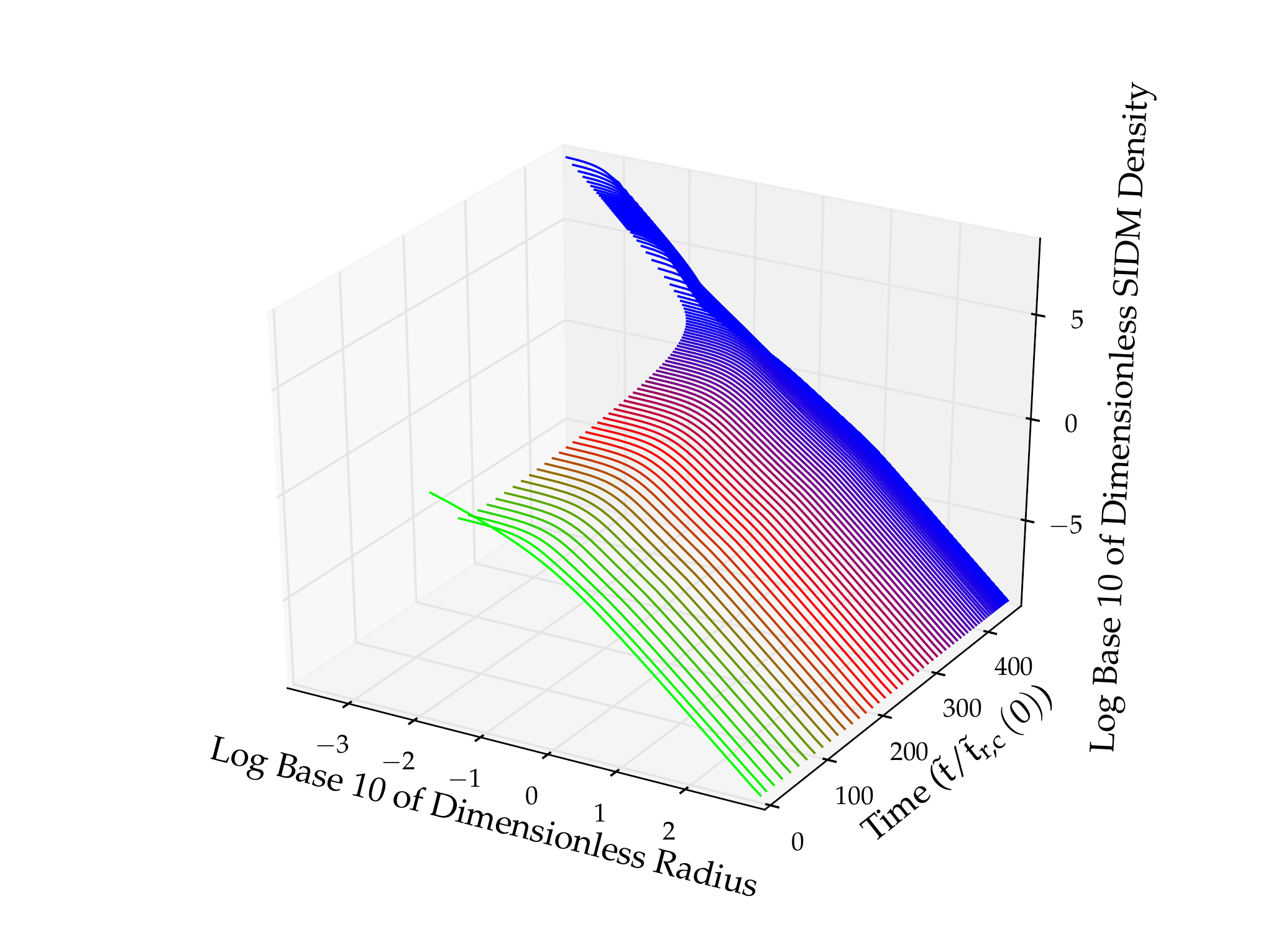}
\caption{Runaway collapse of an $f=1$ SIDM halo with $\hat{\sigma}=0.088$, starting from an initial NFW profile. 
The inner profile starts cuspy, rapidly shrinks to a self-similar profile (as in \cite{Balberg:2002ue} and Figure \ref{fig:initial evolution} above) with a $\tilde{\rho}=1$ core, then slowly increases in density in a self-similar manner. 
After $\sim450$ relaxation times, the core of the halo becomes optically thick, and self-similarity is broken: the core splits into a very dense inner core and an outer core which transitions between the two regions. 
Catastrophic collapse occurs as $\tilde{t}/\tilde{t}_{\mathrm{r,c}}(0)$ approaches $\sim455.65$. 
%As mentioned in subsection \ref{sub:integration} above, the size of successive timesteps decreases rapidly as the gravothermal catastrophe approaches, so on this plot, with a linear time scale, the profiles appear to pile up on top of each other. 
\label{fig:collapse_sfmp}}
\end{figure}

\begin{figure}
\centering{}\includegraphics[width=1\linewidth]{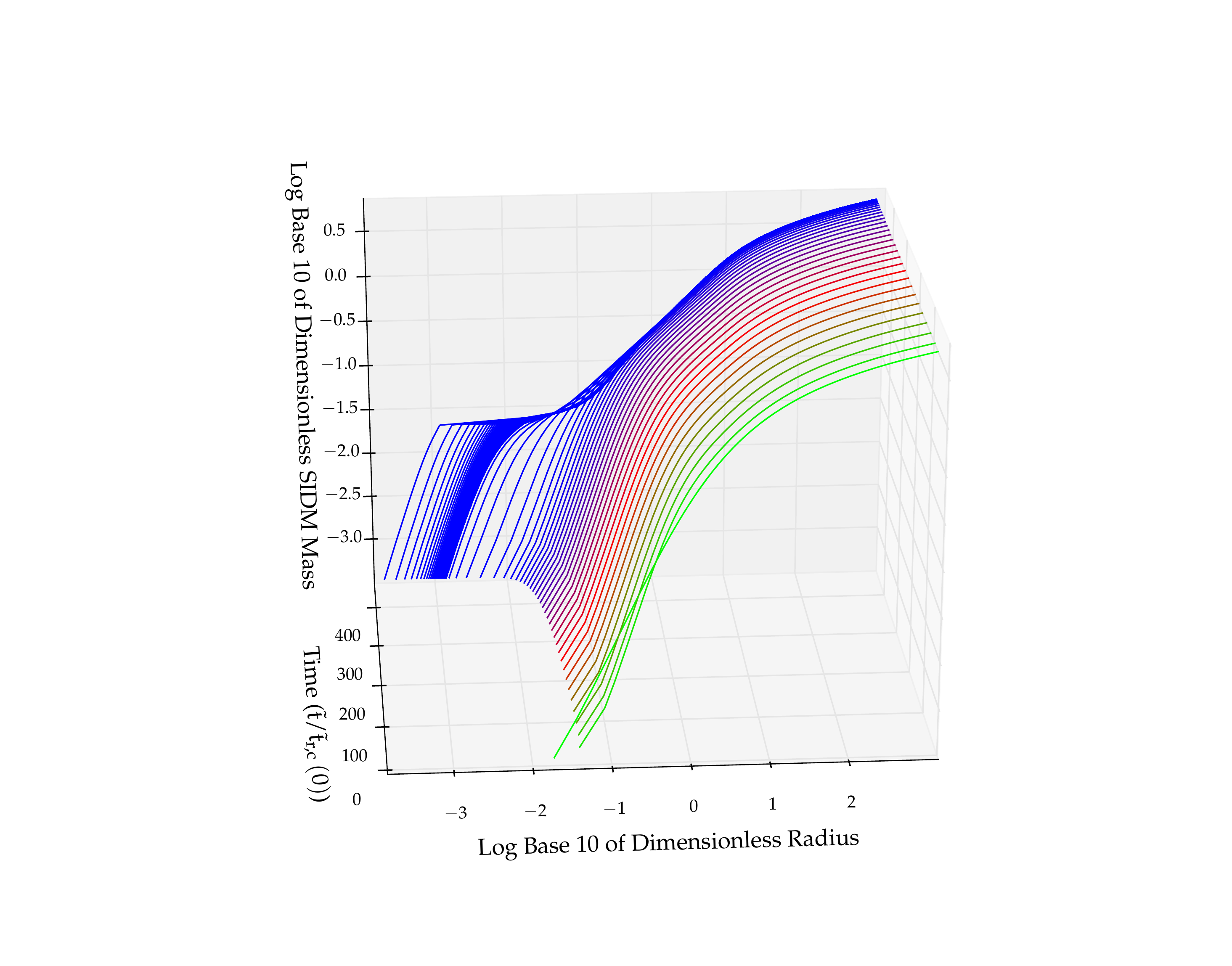}
\caption{Mass profile history of a cored SIDM halo with $\hat{\sigma}=0.088$, starting from an initial NFW profile. 
Once the core enters the optically thick regime, around $\tilde{t}/\tilde{t}_{r,c}(0)=450$, the inner core contains a constant total mass, around $2.5-3\%$ of the characteristic mass $M_{0}$. 
\label{fig:collapse_smfp_mass}}
\end{figure}

We will focus on the $f=1$ case in the following, and then rescale our final results by $f^{-1}$ as just described.
Figures \ref{fig:collapse_sfmp} and \ref{fig:collapse_smfp_mass} show the entire evolution of an $f=1$ halo with $\hat{\sigma}=0.088$ (chosen to allow comparison with \cite{2011MNRAS.415.1125K}) from initial NFW profile through to gravothermal collapse. 
First consider Figure \ref{fig:collapse_sfmp}, which shows the evolution of the density profile.
Although the halo is initially in an NFW profile, the initial negative luminosity at small radii causes the cusp to empty out, driving evolution towards the cored, self-similar profile found by \cite{Balberg:2002ue}, as was already seen in Figure \ref{fig:initial evolution} above. 
When the self-similar profile is reached after a few tens of relaxation times, the luminosity profile becomes everywhere positive, and the core increases in density while its mass steadily shrinks.
While the entire profile is in the lmfp regime, evolution is self-similar, and the central density increases steadily.
Inevitably, there comes a time, about 450 relaxation times after virialization, when the inner density increases enough that the most central regions enter the smfp regime, and the core bifurcates into a very dense outer core and an inner core which transitions between the two regions.

Importantly, once the smfp regime has been reached, mass loss from the inner core is no longer efficient: the inner core has become so thick that evaporation is only possible from its boundary, not from the entire volume. 
This means that the mass in the inner core is essentially constant over the very short time ($\apprle10t_{r,c}(0)$) between breaking of self-similarity and catastrophic collapse.
As mentioned in subsection \ref{sub:integration} above, the size of successive timesteps decreases rapidly as the gravothermal catastrophe approaches, so this short time takes very many (increasingly small) timesteps to integrate over, and the time of collapse can be precisely given as $455.65$ relaxation times after the start of integration. 
Because evaporation is inefficient after the loss of self-similarity, the mass in the inner core is still nonzero at the moment of collapse, unlike in the globular cluster case, and a black hole will form. 
Figure \ref{fig:collapse_smfp_mass} shows that the inner core at collapse contains a mass of around $0.025M_{0}$.
Because the fluid approximation breaks down, we do not know that the entire inner core will collapse directly into a black hole, but, because it is optically thick, Bondi accretion \cite{Bondi:1952ni} is extremely efficient. Hence, we expect that the black hole will rapidly grow to encompass the entire region regardless.

%%%%%%%%%%%%%%%%%%%%%%%%%%%%%%%%%%%%
\section{Supermassive Black Holes from uSIDM}

We have found that halos with a uSIDM component (and pure SIDM halos, on much longer timescales) grow black holes of mass $M_{BH}\equiv0.025fM_{0}$ in a time $455.65t_{r,c}(0)$. 
Given the considerations discussed above, uSIDM can help explain the existence of massive high-redshift quasars if the resulting black holes are large enough and form early enough that baryonic accretion can grow them to $\sim10^{9}M_{\odot}$ by $z\apprge6$. 
It remains to evaluate $M_{0}$ and $t_{r,c}(0)$ in terms of the halo parameters and use this requirement to place constraints on the uSIDM parameters $\{\sigma,\ f\}$.

%%%%%%%%%%%%%%%%%%%%%%%%%%%%%%%%%%%%
\subsection{Halo Parameters}

Instead of using the characteristic NFW parameters $\{\rho_{s},r_{s}\}$, it is convenient to again parameterize a halo by its virial mass $M_{\Delta}$ and concentration $c$. 
In dimensionless units, the mass contained within the $i$th shell is
\begin{equation}
\tilde{M}_{i}=\int_{0}^{\tilde{r_{i}}}\tilde{\rho}\tilde{r}^{2}d\tilde{r}\\
=\int_{0}^{\tilde{r}_{i}}\tilde{r}^{-1}\left(1+\tilde{r}\right)^{-2}\tilde{r}^{2}d\tilde{r}\\
=-\frac{\tilde{r}_{i}}{1+\tilde{r_{i}}}+\ln(1+\tilde{r_{i}}),
\end{equation}
but the virial radius $r_{\Delta}\equiv cr_{s}$, so
\begin{equation}
\frac{M_{BH}}{M_{\Delta}}=\frac{\tilde{M}_{BH}}{\tilde{M}(c)}=\frac{0.025f}{\ln(1+c)-c/(1+c)}.\label{eq:bh mass}
\end{equation}
This gives the desired expression for the seed black hole mass $\tilde{M}_{BH}$ in term of the halo and uSIDM parameters. The denominator ranges from $\sim0.5-2$ for realistic values of the halo concentration, so the BH mass is a few percent of the total uSIDM mass in the halo.

Recall that the relaxation time is $t_{r,c}(0)=1/(af\rho_{s}\nu_{s}\sigma)$, i.e.\ the scattering time at the characteristic radius. 
The lower end of the interesting range for $f=1$ SIDM is $\sim0.1\ \mathrm{cm^{2}/g}$, for which the relaxation time at the characteristic radius of a Milky-Way scale halo is approximately a Hubble time. 
To grow a black hole in galactic halos by $z\sim6$, the relaxation time needs to be $\sim10^{4}$ times smaller to ensure $\sim500$ relaxation times by the time the universe was a twentieth of its present age. 
This does not mean that $\sigma\approx1000f^{-1}\ \mathrm{cm^{2}/g}$, though! 
Recall that $\rho_{s}=\delta_{c}\rho_{crit}$, where $\delta_{c}$ is a function of $c$ and the cosmology given below, and the critical density goes as $(1+z)^{3}$ in the matter-dominated era. 
Also $r_{s}\propto r_{\Delta}\propto(M_{\Delta}/\rho_{crit})^{1/3}$ implies $\nu_{s}\propto\sqrt{M_{\Delta}/r_{\Delta}}\propto M_{\Delta}^{1/3}\rho_{crit}^{1/6}$.
Hence the mass and approximate redshift dependence of the relaxation time are
\begin{equation}
t_{r,c}(0)\propto(1+z)^{-7/2}M_{\Delta}^{-1/3}\label{eq:approximate_collapse_time}
\end{equation}
and we expect that $\sigma f$ need not be that much larger than the interesting range for $f=1$, i.e.\ we expect $\sigma f\apprge 0.1-1\ \mathrm{cm^{2}/g}$.

At this point, the reader might worry that this conclusion combined with the observation of non-collapsed cores in the nearby ($z\sim0$) universe rules out the existence of standard ($f\approx1$) SIDM.
We emphasize, however, that large values of $\sigma f$ mean that core collapse in times much smaller than the age of the universe is \textit{possible}, but not, we expect, \textit{typical}; it occurs only in the rare halos which virialize at very high redshifts and remain uninterrupted, i.e.\ do not experience major mergers, for long enough to complete the gravothermal collapse process. 
%For example, the estimates presented in the example below show that a halo of mass $M_\Delta=10^{12}M_{\odot}$ formed at redshift 15 will undergo core collapse in $\sim161\ \mathrm{Myr}\times(1\ \mathrm{cm^{2}/g})/(\sigma f)$, i.e.\ in $\sim1.61\ \mathrm{Gyr}$ for $\sigma\approx0.1\ \mathrm{cm^{2}/g}$, at the lower end of the interesting range.
%We cannot conclude from this that the Milky Way must have undergone core collapse if the dark matter is self-interacting, though!
%Again, the point is that the expression for $t_{r,c}(0)$ is strongly dependent on the redshift when the collapse process began.
%As we will discuss in the conclusion, we should expect gravothermal collapse to be disrupted by merger activity.
%So it is only the rare halo that did not undergo any major mergers for an extended period of time that undergoes core collapse and grows a black hole this early.
%The typical halo will undergo many mergers during its early history.%
See subsection \ref{sec:caveats} below for further discussion of this point.  

The exact expression for the halo relaxation time is
\begin{equation}
t_{r,c}(0)=\frac{1}{af\sigma}(\frac{K_c^{2}}{4\pi G^{3}})^{1/6}\delta_{c}^{-7/6}\rho_{crit}(z)^{-7/6}M_{\Delta}^{-1/3}\label{eq:exact_collapse_time}
\end{equation}
\begin{equation}
=\mathrm{0.354\ Myr}\times\left(\frac{M_{\Delta}}{10^{12}M_{\odot}}\right)^{-1/3}\left(\frac{K_{c}}{K_{9}}\right)^{3/2}\left(\frac{c}{9}\right)^{-7/2}\left(\frac{\rho_{crit}(z)}{\rho_{crit}(z=15)}\right)^{-7/6}\left(\frac{\sigma f}{1\mathrm{\ cm^2/g}}\right)^{-1},
\end{equation}
where $K_c\equiv\ln(1+c)-c/(1+c)$, $\delta_{c}=(\Delta/3)c^{3}/K_c$, and $\Delta$, the virial overdensity, is $18\pi^2\Omega_{m}^{0.45}$ for a flat universe, approximately 178 in the matter-dominated era.
So the relaxation time, and hence the collapse time, is given in terms of the halo and uSIDM parameters. To match observations, we need some seed black holes to grow by a large enough factor via Eddington accretion to reach $M_{BH}\approx10^{9}M_{\odot}$ by $z\apprge6$; this leads to an inequality on $\sigma$ when the halo parameters and $f$ are specified.

%%%%%%%%%%%%%%%%%%%%%%%%%%%%%%%%%%%%
\subsection{Explaining Observations}

Let us spell out the procedure more precisely.
An observation of a particular high-redshift quasar at redshift $z_{obs}$ yields a value for the luminosity, which corresponds to a supermassive black hole of mass $M_{SMBH}$ once the measured luminosity is identified with the Eddington luminosity and a particular value for the radiative efficiency $\epsilon_r$ is assumed. 
(We have already discussed potential issues with these assumptions in section \ref{sec:smbh} above; in the remainder of the paper, we will take the published observations at face value and assume their quoted SMBH masses, which take $\epsilon_r$=0.1 as input, are correct.)

At the same time, the uSIDM framework developed in this paper tells us that NFW halos of viral mass $M_\Delta$ and concentration $c$ virialized at redshift $z$ form seed black holes of mass $M_{BH}$ in a time $455.65t_{r,c}(0)$, i.e.\ the seed black holes are formed at redshift $z_{coll}$, where 
\begin{equation}
t(z_{coll})-t(z)=455.65t_{r,c}(0),
\label{eq:collapse_time}
\end{equation} 
and the time $t(z)$ after the Big Bang corresponding to redshift $z$ is given by the usual cosmology-dependent expression, 
\begin{equation}
t(z)=t_{0}\int_{0}^{1/(1+z)}\frac{da}{\dot{a}}.
\end{equation}
Equations (\ref{eq:bh mass}) and (\ref{eq:exact_collapse_time}) then give expressions for these quantities in terms of the halo properties $\{M_\Delta,c,z\}$ and the uSIDM parameters $\{\sigma,f\}$.

There is still one parameter that must be specified: the fraction of SMBH mass which is due to accretion of baryons as opposed to the initial seed black hole. 
For simplicity, we will assume that the central black hole accretes continuously at the Eddington limit from the time of formation to the time at which it is observed. 
Of course more complicated growth histories are both possible and likely.
Nevertheless, this simplifying assumption allows us to specify the fraction by instead giving $N_e$, the number of $e$-folds of accretion at the Eddington limit.
This finally allows us to compute the observable quantities: we have 
\begin{equation}
t(z_{obs})=t(z_{coll})+N_e t_{\mathrm{Sal}}\label{eq:z_obs},
\end{equation}
\begin{equation}
M_{SMBH}=M_{BH}\exp(N_e)\label{eq:M_SMBH}.
\end{equation}
To find acceptable values of $\sigma$ and $f$ given the SMBH observables, we must specify (or marginalize over) the halo parameters and the baryonic contribution to the SMBH mass.
The latter quantity directly sets (\ref{eq:z_obs}) the redshift of seed black hole collapse, $z_{coll}$, which yields the required collapse time and thus the required value of $\sigma f$ via (\ref{eq:exact_collapse_time}).
Knowing the growth due to accretion of baryons also tells us (\ref{eq:M_SMBH}) the required seed black hole mass $M_{BH}$, which specifies $f$ via (\ref{eq:bh mass}).

%%%%%%%%%%%%%%%%%%%%%%%%%%%%%%%%%%%%
\subsection{Examples}\label{sub_examples}

%wolfram alpha input:
%((ln(10)-9/10)^2/(4*pi*(G)^3))^1/6 * ((178/3)*9^3/(ln(10)-9/10))^(-7/6) * (3*(67.04 km/sec/Mpc)^2 * (.3183*16^3+(1-.3183))/(8*pi*newton constant))^(-7/6) (10^12 solar masses)^(-1/3) /(1 cm^2/g)
%then divide by sqrt(16/pi)
%Get times from http://www.astro.ucla.edu/~wright/CosmoCalc.html using H0=67.04, omegam=.3183 from Planck.
As an example, consider again ULAS J1120+0641, with mass $M_{SMBH}\approx2\times10^{9}M_{\odot}$ at $z_{obs}=7.085.$ 
To grow four orders of magnitude ($N_e=\ln 10^4$) by Eddington-limited baryon accretion, for example, 
we must form a seed black hole with mass $M_{BH}=2\times10^{5}M_{\odot}$ by $z_{coll}=12.9$. 
With a halo of mass $M_\Delta=10^{12}M_{\odot}$ and concentration $c=9$ formed at redshift $z=15$, we find that $t_{r,c}(0)=0.354\ \mathrm{Myr}\times(1\ \mathrm{cm^{2}/g})/(\sigma f)$.
In order for $455.65$ relaxation times to have passed in the $64.5\ \mathrm{Myr}$ between $z=15$ and $z_{coll}=12.9$, we must have $\sigma f=2.50\ \mathrm{cm^{2}/g}$.
From (\ref{eq:bh mass}), we require $f=1.12\times10^{-5}$ to get the correct seed mass, so $\sigma=2.23\times10^{5}\ \mathrm{cm^{2}/g}=3.97\times10^{5}\ \mathrm{b/GeV}$.
The large value of $\sigma$ is unsurprising: we chose to start with a halo much larger than the seed black hole we wanted to form, so $f$ had to be small and $\sigma$ large in order to compensate. 

Alternatively, we could start with the same halo but produce the black hole entirely from uSIDM. 
The relaxation time is unchanged: $t_{r,c}(0)=0.354\ \mathrm{Myr}\times(1\ \mathrm{cm^{2}/g})/(\sigma f)$. 
But now the black hole need not form until $z=7.085$, $479\ \mathrm{Myr}$ after halo formation, so the required value of $\sigma f$ is smaller, $\sigma f=0.336\ \mathrm{cm^{2}/g}$. 
Again applying (\ref{eq:bh mass}) yields $f=0.112$, $\sigma=2.99\ \mathrm{cm^{2}/g}=5.36\ \mathrm{b/GeV}$, coming much closer to the classic SIDM cross section.

Of course, in the absence of direct measurements of the host halo of ULAS J1120+0641 the problem is underdetermined. 
The point is that $\sigma f$ takes reasonable values of $\mathcal{O}(1)\ \mathrm{cm^{2}/g}$, well within the regime described by the gravothermal fluid approximation starting from an initial NFW profile.

%%%%%%%%%%%%%%%%%%%%%%%%%%%%%%%%%%%%
\section{Discussion}\label{discussion}

The examples in the previous section show that \textit{individual} observations of high-redshift quasars can successfully be explained within the uSIDM paradigm.
For uSIDM to be fruitful, however, we should ideally be able to find (or rule out) a consistent choice of these parameters which successfully explains the \textit{cosmological} abundance of high-redshift quasars.
It is unsurprising that some choice of cross section per unit mass $\sigma$ and fraction $f$ can reproduce one particular observation, e.g.\ ULAS J1120+0641, but it is more suggestive if that choice can reproduce the entire observed number density of supermassive black holes as a function of mass and redshift.
The minimal requirement for a viable uSIDM model is that it explain (or at least not conflict with) what has currently been observed.
That means producing the correct abundance of $\sim 10^9 M_\odot$ quasars at redshift $6-7$, as has already been discussed, and ensuring that supermassive black holes are not overproduced in the nearby (lower-redshift) universe.
Beyond that, one would like to make concrete predictions for the next generation of experiments, which should be sensitive to smaller masses and higher redshifts.

This task is difficult for a number of reasons.
The essential problem is that a number of nuisance parameters must be constrained or marginalized over in order to connect the uSIDM properties to the SMBH distribution (and then further to the quasar distribution). Even in the simplified setup described above there were already the $e$-folds of baryonic accretion, $N_e$, and the halo parameters $M_\Delta$ and $c$. 
In the cosmological context, these nuisance parameters are promoted to entire unknown functions that are currently only poorly constrained by observations and simulations.
Even when constraints or functional forms are available, they are often trustworthy only in regimes far separated from the ones of interest to us here (for example, in the low-redshift universe, or in a lower mass range).
We have already encountered this problem in Section \ref{sec:gravothermal_collapse} above, when considering the concentrations of massive NFW profiles at high redshifts.

Nevertheless, in the remainder of this section we attempt to estimate the constraints that our existing knowledge places on the uSIDM parameter space.
We first explain the source of our cosmological uncertainty and means by which it could be improved.
Next we note a different source of tension within $\Lambda\mathrm{CDM}$, independent of the existence of high-redshift SMBHs, that could be relieved by uSIDM.
Finally, we present tentative maps of the uSIDM parameter space relevant to the resolution of these tensions.

\subsection{Cosmological Caveats}\label{sec:caveats}

Predicting the cosmological consequences of gravothermal collapse given a choice of the uSIDM parameters requires a unified picture of the SIDM profile at galaxy formation in terms of the halo mass and redshift, which will be easier given proper $N$-body simulations of halos containing uSIDM. 
There are several reasons why using the fluid approximation to simulate an isolated halo does not suffice.

First, although the process of gravothermal collapse can be quite short on cosmological timescales, which is why it allows massive quasars to form faster than in the standard $\Lambda\mathrm{CDM}$ picture, we have seen that it is long in terms of halo time scales (several hundred characteristic relaxation times). 
It is therefore necessary for the halo to remain essentially undisturbed for this length of time in order for core collapse to occur and seed black holes to form..
The beginning and end of the collapse process---the elimination of the initial cusp and the catastrophic collapse itself after the core becomes optically thick---are driven entirely by dynamics in the innermost part of the profile, so we might expect them to be insensitive to accretion or mergers in the outer halo. 
Figures \ref{fig:initial evolution} and \ref{fig:collapse_sfmp} make clear, though, that these stages are very short compared to the length of the overall process.
The vast majority of the time required for collapse involves the slow increase of density in the core as mass flows inward from the outer halo, which we expect to be sensitive to accretion or mergers.
In other words, the halo must be isolated for several hundred relaxation times.
Strong interactions with other masses, such as major mergers, will disrupt the collapse process, essentially resetting the clock for seed black hole formation.
Even accounting for more controlled accretion via minor mergers will technically necessitate the tracking of substructure within the collapsing halo, since it breaks the spherical symmetry required by the fluid approximation, although we expect it will not change our qualitative conclusions. Such tracking of substructure is only truly possible using $N$-body simulations.

More importantly, determining how often the collapse process is disrupted, and therefore predicting the spectrum of black hole masses as a function of redshift for particular values of $\sigma$ and $f$, in order to compare with existing and upcoming observations, requires detailed cosmological information.
We need not only the halo mass function at very high redshifts (up to redshift $15$ in the above example, and ideally out to at least $z\apprge30-50$) but also information on halo shape (the concentration parameter $c(M_\Delta,z)$ at the same high values of z, in the case that the halos form in NFW profiles) and, most importantly, detailed merger probabilities and histories as functions of mass and redshift.
Even when analytical approximations to these quantities at $z\apprle1$ exist, it is unclear how confidently they can be extrapolated to $z\sim50$. Hence dedicated $N$-body simulations are desirable. 
We will briefly note some additional interesting results, beyond the prediction of the history of the black hole mass function, which could be investigated given this cosmological information.

\subsection{The Too Big to Fail Problem}

This paper has noted that gravothermal collapse of uSIDM can produce seed black holes in the center of virialized halos.
We have primarily been concerned with using this mechanism to explain the abundance of massive high-redshift quasars, but we now mention a few other areas where it could prove useful.
We emphasize that these are logically independent of the quasar issue: we should not necessarily expect that the same choice of uSIDM parameters will be useful in both cases.

First, it is intriguing that there exists a well-known (and relatively tight) relation between the properties of a host galaxy and the massive black hole it contains, the $M$--$\sigma$ relation \cite{Magorrian:1997hw,Ferrarese:2000se,Gebhardt:2000fk}, which suggests some sort of causal mechanism connecting the central portions of the galaxy containing the black hole with the more distant regions where the velocity dispersion is measured.
Gravothermal collapse naturally provides one such mechanism, and it would be suggestive if it produced the correct relation for some choice of the uSIDM parameters.
At a minimum, it should not spoil the observed relationship in nearby galaxies; this has been used previously to constrain the cross section of $f=1$ SIDM \cite{Hennawi:2001be,Hu:2005cd}.

More speculatively, the presence of central black holes in dwarf galaxies could resolve the ``too big to fail" problem \cite{BoylanKolchin:2011de, BoylanKolchin:2011dk}, in which the central densities of the brightest Milky Way satellites have much lower central densities than the most massive subhalos in $\Lambda\mathrm{CDM}$ simulations of Milky-Way sized galaxies. 
One way to resolve the problem is to invoke physics not present in the simulations to reduce the central densities (within $\sim1\ \mathrm{kpc}$ of the subhalo center) by a factor of order unity. 
If all of the dark matter is self-interacting with $\sigma\simeq0.1\ \mathrm{cm^{2}/g}$, it naturally smooths out cusps to form cores, which could provide the needed reduction in density \cite{Rocha:2012jg}.
But the small fractions $f\ll1$ we consider in this paper cannot solve the problem in this manner; another method of removing substantial mass from the central $\sim\mathrm{kpc}$ is needed.

Under some circumstances, it is possible that black holes could provide the needed reduction in mass.
Merging black hole binaries emit gravitational waves anistropically and thus receive an impulsive kick, up to several hundred $\mathrm{km/s}$. 
This energy can be distributed to the surrounding baryons and kick out a substantial portion of the central mass, forming a core \cite{Merritt:2004xa, BoylanKolchin:2004tf, Lippai:2008fx}.
Such a scenario is only viable if the required binary black hole mergers are sufficiently common within dwarf galaxies or their progenitors.
Although the standard cosmological model predicts the presence of black holes in the center of nearly all large halos, it is not clear that $\Lambda\mathrm{CDM}$ produces enough black holes within the smaller halos which are the progenitors of dwarf galaxies.
Here we propose instead to use uSIDM to produce them.

Solving the Too Big to Fail problem using black holes formed from uSIDM requires a particular sequence of events: first, small halos must remain isolated enough to form seed black holes; second, the probability of major mergers must become large enough that essentially all of the Milky Way satellites have binary black holes coalesce within them in order to reduce their central densities. 
During the epoch of matter domination, we see that the black hole formation time for a halo of fixed mass goes as $(1+z)^{-7/2}$ (\ref{eq:approximate_collapse_time}), while we expect the merger timescale to be set roughly by the Hubble time, $H^{-1}(z)\sim(1+z)^{-3/2}$. 
So halos of a given mass that form before some critical redshift will indeed grow black holes before they merge.
In the next subsection we consider the parameter space where black hole seeds are ubiquitously formed in the progenitors of today's dwarf galaxies.

\subsection{Parameter Space}

\subsubsection{High-Redshift Quasars}

In subsection \ref{sub_examples} above, we presented two possible routes to produce a supermassive black hole matching observations. 
Here we move from specific examples to a discussion of the entire parameter space relevant to the production of high-redshift quasars like ULAS J1120+0641.
Recall that we have six input parameters: $\{M_\Delta,c,z,N_e,\sigma,f\}$, respectively the halo mass, concentration, redshift of virialization, $e$-folds of Eddington-limited accretion after collapse, and uSIDM cross section per unit mass and fraction.
We specify the halo properties as above: $M_\Delta=10^{12} M_\odot$, $c=9$, $z=15$.
We then use (\ref{eq:z_obs}) and the redshift at which a quasar is observed, in this case $z_{obs}=7.085$, to eliminate $N_e$, leaving a two-dimensional parameter space for production of black holes by this time.
Finally, the requirement that the mass of the quasar match observations, $M_{SMBH}\approx2\times10^{9}M_{\odot}$, combined with the assumption of continuous Eddington-limited growth since black hole formation, reduces the parameter space to one dimension, a curve $\sigma(f)$.

\begin{figure}
\centering{}\includegraphics[width=1\linewidth]{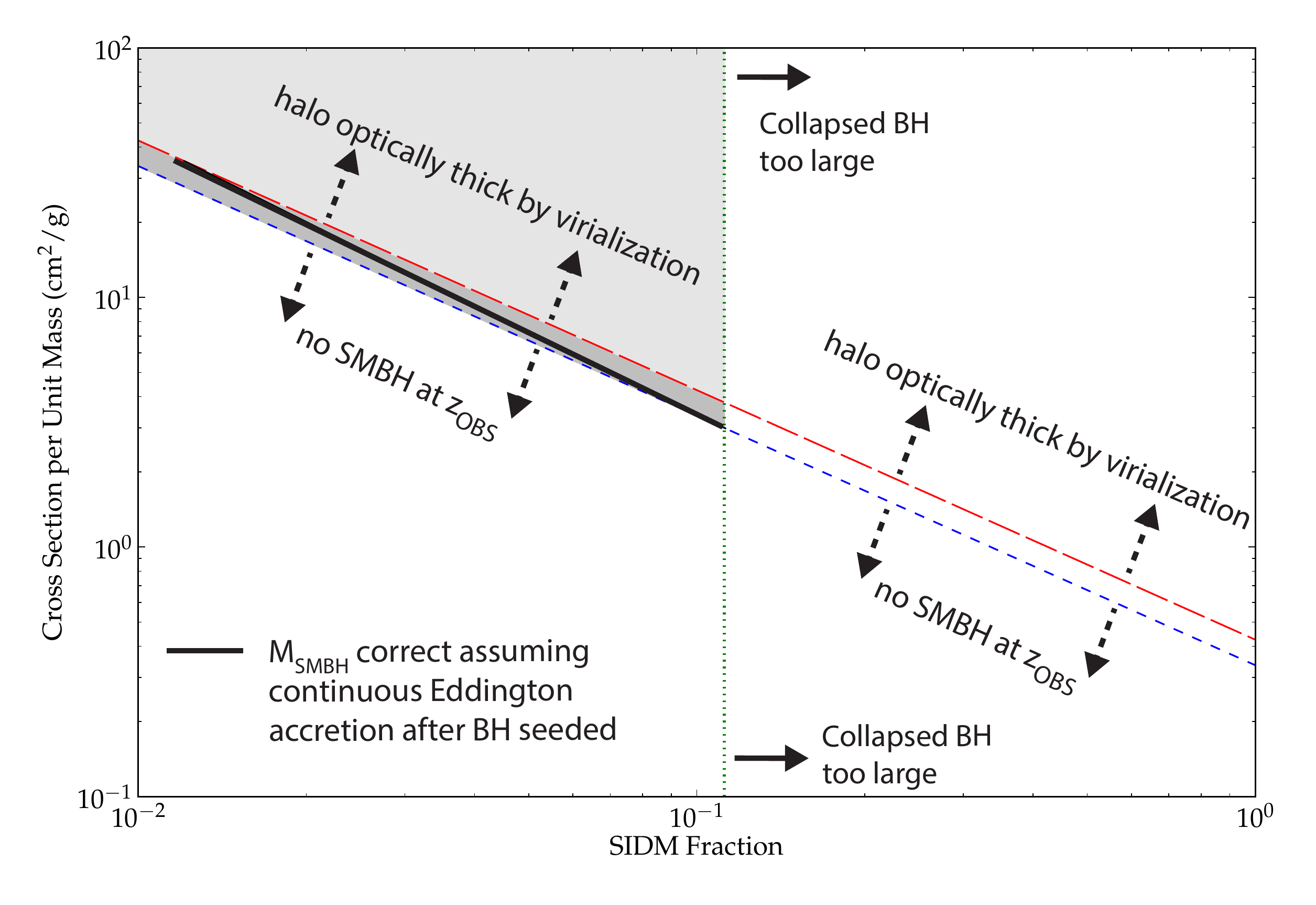}
\caption{uSIDM parameter space for production of massive high-redshift quasars. 
We have used the numbers considered in the example above: $M_{SMBH}\approx2\times10^{9}M_{\odot}$, $z_{obs}=7.085$, $M_\Delta=10^{12}M_{\odot}$, $c=9$, $z=15$. 
The solid line plots values of $\sigma$ and $f$ that result in an SMBH of the desired size at the time of observation, assuming continuous Eddington accretion from the time the core collapses and the seed black hole is formed. 
The green dotted vertical line marks the largest allowed value of $f$. 
To its right, collapsed black holes are already larger than $M_{SMBH}$.
To its left, collapsed black holes form smaller than $M_{SMBH}$, but can grow larger by accreting baryons.
The points on the blue dashed line all result in collapse precisely at the redshift of observation; below this line, a black hole has not yet formed by $z_{obs}$.
Points on and above the red dashed line result in a halo that is already optically thick at the time of virialization, i.e.\ optically thick at the characteristic radius for the initial NFW profile.
As discussed in the text, the methods used in this paper are not directly applicable here, but we still expect gravothermal collapse.
Numerical values for all of these bounding lines are given in the text.
\label{fig:quasarparam}}
\end{figure}

We present the parameter space in Figure \ref{fig:quasarparam}.
%We show the one-dimensional curve $\sigma(f)$, but in light of the uncertainties about baryonic accretion discussed in section \ref{sec:smbh} we also relax the assumption of continuous Eddington-limited growth and show the region of the full $\sigma$--$f$ plane in which black holes are produced.
The one-dimensional curve $\sigma(f)$, where continual Eddington-limited accretion since black hole formation results in a supermassive black hole with $M_{SMBH}=2\times10^{9}M_{\odot}$ at $z_{obs}=7.085$, is the solid black line.
Because the baryonic accretion history after seed black hole formation is uncertain, as discussed in Section \ref{sec:smbh} above, we also indicate with the shaded regions the entire portion of the full $\sigma$--$f$ plane in which black holes of any size smaller than $M_{SMBH}$ are produced by $z_{obs}$.

There are several constraints on this reduced parameter space.
First is the simple requirement that gravothermal collapse indeed occurs before $z_{obs}=7.085$.
We have already seen in subsection \ref{sub_examples} above that this constrains $\sigma f\ge0.336\ \mathrm{cm^{2}/g}$.
Second is the requirement that the black hole produced by gravothermal collapse must not be larger than the observed mass of ULAS J1120+0641.
Combined with additional assumptions about baryonic accretion, this excludes the entire region above the black curve in Figure \ref{fig:quasarparam}.
Even without the assumptions, this still constrains the black hole mass via equation (\ref{eq:bh mass}), and therefore the uSIDM fraction $f$, provided that a black hole is actually produced.
Again, the resulting constraint was calculated in subsection \ref{sub_examples}: $f\le0.112$. 
Larger values of $f$ would produce black holes which contained too large a portion of the mass of the entire halo.

Finally, recall that our expressions for the collapse time and resulting black hole mass are based on simulations.
As discussed in subsection \ref{sub:initial} above, the simulations assume the uSIDM is initially in NFW profile, which is only valid if uSIDM interactions were slow compared to the timescale of halo formation, i.e.\ when the halo is initially optically thin.
This places a constraint on $\sigma f$ as a function of the halo parameters, given by equation (\ref{optically_thin_bound}).
For our chosen values this gives $\sigma f\le0.425\ \mathrm{cm^{2}/g}$.

Because $\sigma f$ directly sets the collapse time via (\ref{eq:exact_collapse_time}, \ref{eq:collapse_time}), the upper bound on $\sigma f$ is also a lower bound on the time of formation of a black hole from an initially optically thin uSIDM halo: in this case, we must have $z_{coll}\le7.90$.
In turn, this places an upper bound on the number of $e$-folds of growth from baryons that can occur before $z_{obs}=7.085$, via (\ref{eq:z_obs}).
We find $N_e\le2.24$, i.e.\ black holes formed from optically thin uSIDM halos have time to grow less than an order of magnitude from baryons.
In particular, we cannot trust the precise results of our simulations for the example we considered in subsection \ref{sub_examples} above, with $N_e=\ln 10^4$.
Note, however, that the upper bound on $\sigma f$, and thus on $z_{coll}$, is independent of $z_{obs}$: it depends only on $z$, the redshift of halo formation.
Most high-redshift quasars are seen near $z_{obs}\sim6$: ULAS J1120+0641 is an outlier.
Black holes at redshift 6 have had time for another $180\ \mathrm{Myr}\sim4t_\mathrm{sal}$ of baryonic growth, so they can grow by up to a factor of $\sim500$ from baryons.

We have seen that there is an extremely narrow range, $0.336\ \mathrm{cm^{2}/g}\le\sigma f\le0.425\ \mathrm{cm^{2}/g}$, in which the uSIDM halo considered here is optically thin at virialization but nevertheless rapidly collapses to form a black hole.
%Inspection of (\ref{eq:opt_thin}) and (\ref{eq:exact_collapse_time}) shows that both have the same mass dependence, but different dependences on redshift and concentration, so for generic values of the halo parameters the values of the two bounds will be quite different.
%The near-coincidence of the two bounds thus requires some explanation.
%Heuristically speaking, we might understand the closeness of these upper and lower bounds in the following way.
%The upper bound requires that the initial halo be optically thin, i.e.\ that the scattering cross section be less than the ``characteristic cross section" of the halo, $1/(\rho_s r_s)$.
%But the lower bound requires that collapse not take too long, i.e.\ that the scattering time at the characteristic radius is small compared to a Hubble time.
%But these bounds are not independent. 
How can we explain the closeness of these two bounds? 
In general, they are not independent.
The upper bound requires that the initial halo be optically thin, i.e.\ that the scattering cross section be less than the ``characteristic cross section" of the halo, $1/(\rho_s r_s)$.
But the lower bound requires that collapse not take too long, i.e.\ that the scattering time at the characteristic radius, $1/(\sigma f \rho_s \nu_s)$ is small compared to a Hubble time.
These bounds can be simultaneously satisfied when $H^{-1}\sim r_s/\nu_s$.
But, ignoring concentration dependence and numerical factors, $\nu_s\sim\sqrt{GM_\Delta/r_s}\sim\sqrt{G \rho_{crit} r_s^2}\sim H r_s$, so $r_s/\nu_s\sim H^{-1}$ as desired. 
That is, both bounds exhibit the same mass dependence, and their redshift dependence is identical when $z$ and $z_{coll}$ are similar, as can be verified from (\ref{eq:opt_thin}, \ref{eq:exact_collapse_time}, \ref{eq:collapse_time}). 
For our particular choice of $c=9$, the numerical factors are nearly canceled by the concentration dependence, so the bounds are especially close.

We emphasize again, however, that the upper bound on $\sigma f$ (the red dashed curve in Figure \ref{fig:quasarparam}) is not a true physical exclusion of the uSIDM parameter space above it.
It merely signals that the fluid approximation used in this paper is no longer valid outside this space.
As discussed extensively in subsection \ref{sub:initial} above, we expect that profiles in which the uSIDM starts optically thick should in fact undergo collapse even faster.
The requirement of an optically thin initial profile would only be physical if starting otherwise led to fragmentation, turbulence, or some other mechanism by which the core was destroyed or core collapse avoided.

Figure \ref{fig:quasarparam} presents the uSIDM parameter space for a particular choice of halo parameters $\{M_\Delta,c,z\}$.
We briefly consider how constraints on the parameter space are changed when these parameters are altered.
First consider the halo mass $M_\Delta$. 
The collapse time (\ref{eq:exact_collapse_time}) scales as $M_\Delta^{-1/3}$, so smaller values of the halo mass require values of $\sigma f$ to form black holes in the same time.
At the same time, the value of $\sigma f$ required for the halo to start initially optically thin (\ref{eq:opt_thin}) has the same scaling with halo mass.
So decreasing $M_\Delta$ will shift both bounds to higher values of $\sigma f$ (up and to the right on the $\sigma$--$f$ plane), but it will not \textit{qualitatively} change the shape of the allowed parameter space.
This shift accounts for the main difference between the high-redshift quasar parameter space and the dwarf satellite parameter space we will consider next.
We note, however, that at relatively low redshifts there is a well-known black hole--bulge relation \cite{Magorrian:1997hw,Marconi:2003hj,Haring:2004hr}, $M_{SMBH}\sim10^{-3}M_{bulge}$.
If this relation persists at high redshifts, we should not depart too far from $M_\Delta\sim10^{12}M_\odot$ to explain $M_{SMBH}\sim10^9 M_\odot$. 

Next consider the concentration parameter $c$.
Again consulting (\ref{eq:exact_collapse_time}), we see that the collapse time depends strongly on concentration, scaling roughly as $c^{-7/2}$.
The collapse time depends more strongly on the concentration than does the optically thin condition, so for small enough values of $c$ it will be impossible to form black holes before a given redshift starting from an optically thin halo.
When the other halo parameters and $z_{obs}$ are kept fixed, we find that this critical value of $c$ is $7.4$.
Conversely, by going to larger and larger values of $c$ we can form black holes by any desired time at smaller and smaller values of $\sigma f$.
However, extremely high values of the concentration parameter correspond (unsurprisingly) to extremely concentrated halos, with $r_s\ll r_\Delta$. 
It is not clear that such halos are actually produced in $\Lambda \mathrm{CDM}$.

Finally, consider the redshift of halo formation $z$.
Once more consulting (\ref{eq:exact_collapse_time}), we see that we can take the collapse time to zero by increasing $z$.
Heuristically, this is because the critical density, and thus characteristic density, increases with increasing redshift, so a just-virialized halo is closer to the densities needed to start the catastrophic collapse process.
However, producing large virialized halos at higher and higher redshifts becomes increasingly unphysical given the bottom-up structure formation mechanism in $\Lambda \mathrm{CDM}$.
Decreasing $z$ has the opposite effect: halos of a given size become more common, but larger values of $\sigma f$ are required to produce a black hole by a given $z_{obs}$.
Like the case of small concentration parameter, for small enough $z$ it is impossible to form black holes starting from optically thin halos before a given time.
In this case the bound on the redshift of formation for the halo considered here is $z>13.53$.

\subsubsection{Dwarf Galaxies}

Recall from the previous subsection that one resolution to the Too Big to Fail problem is the formation of cores in dwarf galaxies if matter is ejected during binary black hole mergers.
Our goal here is to specify the parameter space in which uSIDM produces black holes in the progenitors of dwarf galaxies before the epoch in which binary mergers are common.
As in the case of high-redshift quasars above, we start by specifying a set of typical values for the halo parameters $\{M_\Delta,c,z\}$.
Ref. \cite{BoylanKolchin:2011dk} compared the Milky Way dwarf galaxies to subhalos around similarly-sized galaxies in the Aquarius simulations \cite{Springel:2008cc} to derive probable values for the virial mass $M_\Delta$ and maximum central velocity $v_\mathrm{max}$ of each halo at the time of its infall into the main Milky Way halo.

Recall that the maximum velocity of an NFW profile is
\begin{equation}
v_\mathrm{max}=0.465\sqrt{\frac{c}{K_c}}v_\Delta,
\end{equation}
with $K_c=\ln(1+c)-c/(1+c)$, at radius
\begin{equation}
r_\mathrm{max}=2.163r_s,
\end{equation}
as can easily be verified numerically using the definition of the NFW density profile (\ref{eq:nfw profile}) and $v=\sqrt{G M(r)/r}$.
Rearranging gives an expression for $K_c/c$ in terms of $M_\Delta$, $v_\mathrm{max}$, and $z_{infall}$. 
In particular, since $r_\Delta\sim\rho_s\sim\rho_{crit}(z)^{-1/3}$, we have $K_c/c\sim\rho_{crit}(z)^{1/3}$.
But $K_c/c$ has a maximum value of $0.216$, so above some value of $z_{infall}$, there is no possible NFW profile with the given values of $M_\Delta$ and $v_\mathrm{max}$. 
For the derived values for the Milky Way dwarfs, we find in general that $z_{infall}\apprle6$. 
As the infall redshift moves lower for each particular dwarf, the concentration increases from a minimal value $c=2.16$. (Actually, $K_c/c$ attains its maximum at $c=2.16$ and approaches zero both as $c\rightarrow0$ and $c\rightarrow\infty$, but we neglect the former branch, with $c<2.16$, as unphysical.)
In particular, we choose $z_{infall}\sim4.5$, which results in $c\sim9$ for a typical dwarf, the same as considered above.

A typical Milky Way dwarf in \cite{BoylanKolchin:2011dk} has $M_\Delta=2\times10^8 M_\odot$ at the time of infall.
If Too Big to Fail is to be explained by means of binary black hole mergers, a typical dwarf should have undergone a major merger, so that a binary black hole merger occurs in the first place.
We will therefore take $M_\Delta=10^8 M_\odot$, $c=9$, $z_{obs}=4.5$ as our typical parameters. 
%
%We still have a choice of $z$, the redshift of progenitor formation.
%We first pick an arbitrary value---$z=15$ once again---in order to plot the parameter space. 
%Below we will consider whether both high-redshift quasars and black holes in dwarf galaxies can both be produced with identical choices of uSIDM parameters, i.e.\ with the same species of self-interacting dark matter.

\begin{figure}
\centering{}\includegraphics[width=1\linewidth]{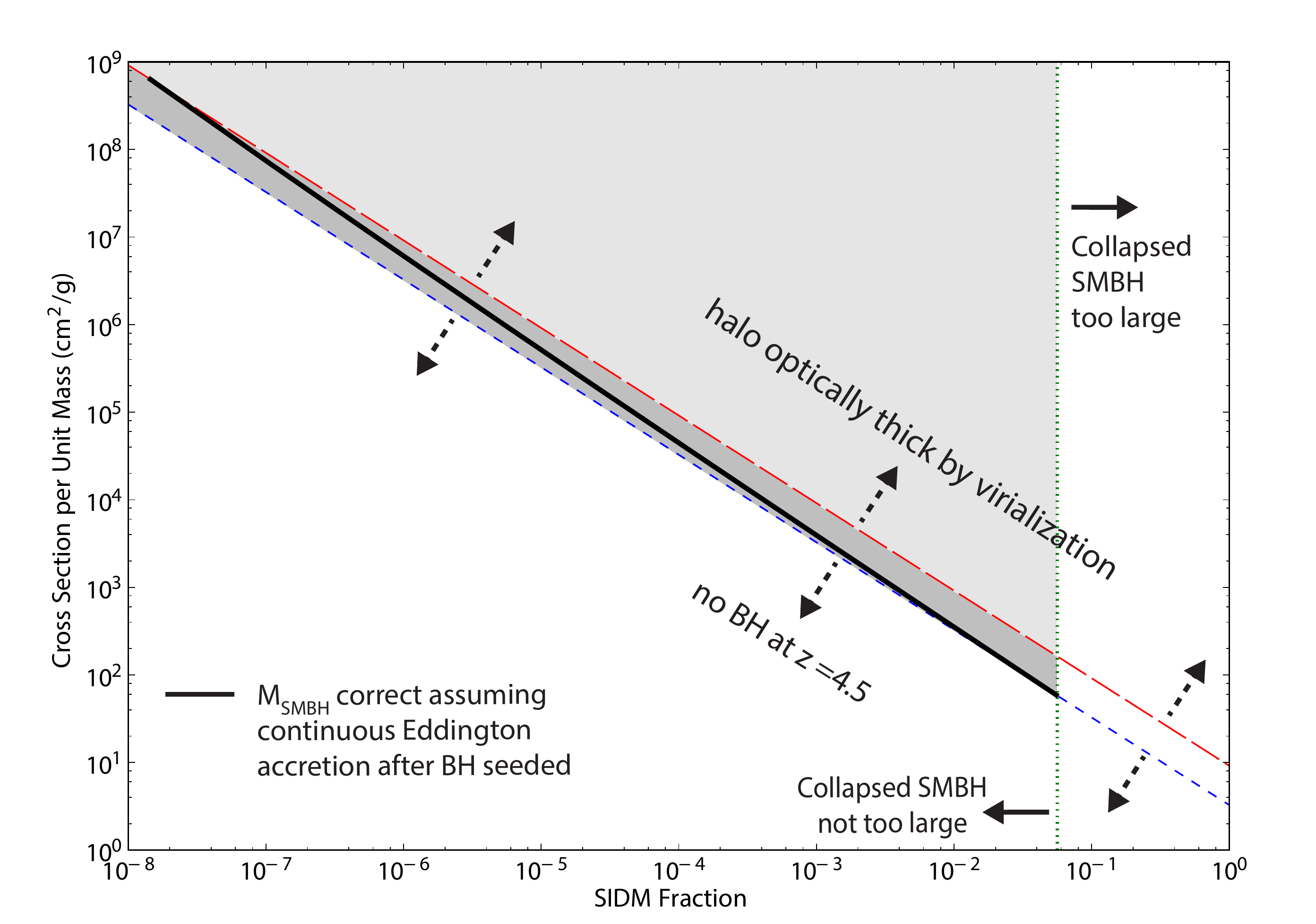}
\caption{uSIDM parameter space for production of black holes in dwarf galaxies. 
The parameters used are $M_{SMBH}=10^5M_{\odot}$, $z_{obs}=4.5$, $M_\Delta=10^{8}M_{\odot}$, $c=9$, $z=15$. 
The solid line plots values of $\sigma$ and $f$ that result in an SMBH of the desired size at the time of observation, assuming continuous Eddington accretion from the time the core collapses and the seed black hole is formed. 
The green dotted vertical line marks the largest allowed value of $f$. 
To its right, collapsed black holes are already larger than $M_{SMBH}$.
To its left, collapsed black holes form smaller than $M_{SMBH}$, but can grow larger by accreting baryons.
The points on the blue dashed line all result in collapse precisely at the redshift of observation; below this line, a black hole has not yet formed by $z_{obs}$.
Points on and above the red dashed line result in a halo that is already optically thick at the time of virialization, i.e.\ optically thick at the characteristic radius for the initial NFW profile.
As discussed in the text, the methods used in this paper are not directly applicable here, but we still expect gravothermal collapse.
Numerical values for all of these bounding lines are given in the text.
\label{fig:dwarfparam}}
\end{figure}

Figure \ref{fig:dwarfparam} presents the parameter space for our typical dwarf halo.
We have taken the black hole mass to be $M_{SMBH}=10^5 M_\odot$, in accordance with the black hole--bulge relation, and again assumed that the redshift of formation of the progenitor halo is $z=15$.
The various bounds in the figure are attained in the same manner as they were for the quasar bounds shown in Figure \ref{fig:quasarparam}, so we simply quote them here and refer to the discussion above for details of their calculation.
The lower bound on $\sigma f$, which comes from requiring collapse before $z_{obs}=4.5$, is $\sigma f\ge3.26\ \mathrm{cm^{2}/g}$.
The upper bound on $f$, which is calculated using (\ref{eq:bh mass}) and scales linearly with $M_{SMBH}$, is $f\le0.056$.

The upper bound on $\sigma f$, set by the requirement of an optically thin initial profile, is $\sigma f\le9.16\ \mathrm{cm^{2}/g}$.
This corresponds to an upper bound on the redshift of collapse, $z_{coll}\le7.90$.
(As discussed above, the upper and lower bounds have the same mass dependence, and we are considering the same values of $z$ and $c$ as we did for high-redshift quasars, so we recover the same bound on the collapse time.)
Once again, this gives an upper bound on the number of $e$-folds of growth from baryons, via (\ref{eq:z_obs}).
Because we have a much longer time for growth after black hole formation than we did in the high-redshift quasar case, it is much looser: $N_e\le15.2$.
Once again, the allowed range on $\sigma f$ is a factor of only a few.
But the significantly looser bound on $N_e$ means that black holes can grow by a factor of nearly $4\times10^6$.
So $f$ can be decreased by over six orders of magnitude from its maximal value, and $\sigma$ increased by a corresponding amount, while still maintaining an optically thin initial profile and allowing reasonably large black holes to form.
This explains the much larger range in $\sigma$ and $f$ seen on Figure \ref{fig:dwarfparam} compared to Figure \ref{fig:quasarparam}.

\subsubsection{Both Simultaneously?}

We have just seen that the minimum value of $\sigma f$ needed to produce massive high-redshift quasars is about an order of magnitude lower than those needed to produce black holes in dwarf galaxies before major mergers.
We can understand this qualitatively from the expression for the halo relaxation time, (\ref{eq:exact_collapse_time}): it scales as $M_\Delta^{-1/3}$.
The black holes in dwarf galaxies have about twice as long to form, until $z_{obs}=4.5$ instead of $z_{obs}=7.085$, so $\sigma f$ is scaled by a factor of $10^{4/3}/2\approx10$.

%If we ignore the requirement of an optically thin initial profile, and assume that there is no functional upper bound on the values of $\sigma f$ that can yield a black hole, then the uSIDM parameters which produce black holes in dwarfs are a strict subset of those which produce high-redshift quasars. 
This scaling of $\sigma f$ implies that the uSIDM parameters which produce black holes in dwarfs are a strict subset of those which produce high-redshift quasars.
It is then easy to choose values which solve both: one simply takes $\sigma f\ge3.26\ \mathrm{cm^{2}/g}$ and chooses compatible values of $\sigma$ and $f$ to taste. 
Because $\sigma f$ is significantly larger than the minimum value needed to produce high-redshift quasars, the uSIDM halos which produce them will start initially optically thick, above the (red dashed) upper bound on $\sigma f$ shown in Figure \ref{fig:quasarparam}.
In this optically thick regime, the expressions for the gravothermal collapse time and black hole seed mass derived from the simulations of Section \ref{sec:gravothermal_collapse} should be taken as limits: we expect that gravothermal collapse should occur a slightly shorter time after halo formation and result in slightly larger seed black holes.

As an example, consider the $\{\sigma,f\}$ values that fall in the one-dimensional parameter space discussed at the beginning of this subsection, where continuous Eddington accretion from the time of black hole collapse until $z_{obs}$ just produces a supermassive black hole with the observed value of $M_{SMBH}=2\times10^9M_\odot$ (the solid black line in Figure \ref{fig:quasarparam}). 
If $\sigma f=3.26$, the smallest possible value needed to also produce black holes in dwarfs by redshift $4.5$, we would find using (\ref{eq:exact_collapse_time}, \ref{eq:z_obs}) that $z_{coll}=13.3$.
(Strictly speaking (\ref{eq:z_obs}) is not valid in the context of an optically thick initial halo, since it uses an expression for the collapse time derived from an initially optically thin NFW profile.
We are simply using it here for the sake of illustration.)
In this case there is time for $9.54$ $e$-folds of baryonic accretion before $z=7.085$, and the initial black hole has mass $7\times10^4 M_\odot$.
This determines the USDIM parameters for this example as $\sigma=8.14\times10^5\ \mathrm{cm^2/g}$, $f=4.01\times10^{-6}$.

As the beginning of this section emphasized, it is largely beyond this scope of this paper to describe a fully consistent uSIDM cosmology.
Nevertheless, this subsection suggests that the USDIM paradigm is flexible enough to resolve both of the potential tensions within $\mathrm{\Lambda CDM}$ discussed here.
It is possible that a single species of ultra-strongly self-interacting dark matter could in fact resolve both tensions simultaneously.
In investigating this question further, it will be important to move beyond the simplifying assumptions employed herein, especially the stipulations of an initial optically thin profile and a cosmologically isolated profile.

\section{Conclusion}

%As promised, we have seen that there exists an interesting portion of the SIDM parameter space where the SIDM component is subdominant but ultra-strongly self-interacting, with $f\ll1$ and $\sigma\gg1\ \mathrm{cm^{2}/g}$. 
In this paper, we considered a minimal extension of the SIDM parameter space, in which a self-interacting component comprises only a fraction of the dark matter.
For $f\apprle0.1$, this evades all prior constraints on SIDM models.
We highlighted the uSIDM regime, where the SIDM component is subdominant but ultra-strongly self-interacting, with $f\ll1$ and $\sigma\gg1\ \mathrm{cm^{2}/g}$. 
In the setup considered here, the presence of uSIDM leads to the production of black holes with a mass of around $2\%$ of the total uSIDM mass in the halo at very early times. 
In particular, such black holes can act as seeds for baryon accretion starting soon after halo formation, alleviating potential difficulties with accommodating massive quasars at high redshifts within the standard $\Lambda\mathrm{CDM}$ cosmology. 
If black holes are formed ubiquitously in dwarf halos before they undergo mergers, they may also resolve the Too Big to Fail problem by ejecting matter from cores during black hole mergers.
More detailed cosmological simulations are needed to confirm the conclusions of this paper and suggest other potential observational consequences of uSIDM.

Setting aside the detailed predictions, this paper has demonstrated that multi-component dark matter can have strong effects on small scales while still evading existing constraints.
In the toy model discussed here, the strong effect was the result of the gravothermal catastrophe.
Gravothermal collapse of a strongly-interacting dark matter component is a novel mechanism for production of seed black holes, potentially one with many implications.
Given its appearance in the simple extension of $\Lambda\mathrm{CDM}$ considered here, it is plausible that gravothermal collapse and its observational consequences, such as seed black hole formation, are generic features of more detailed models.
It is important to consider, and then observe or constrain, this and other observational consequences that are qualitatively different from the predictions of the standard cosmological model.

Our discussion has been purely phenomenological, so it is reassuring to note the existence of a class of hidden-sector models \cite{Boddy:2014yra} which naturally produce a subdominant strongly-interacting dark matter component, with self-interaction cross-sections ranging as high as $\sigma\sim10^{11}\ \mathrm{cm^{2}/g}$.
Very interestingly, some models give both a dominant component with $\sigma\simeq0.1-1\ \mathrm{cm^{2}/g}$, as needed to alleviate discrepancies between $\Lambda\mathrm{CDM}$ and observations, and a uSIDM component with $\sigma\simeq10^5-10^7\ \mathrm{cm^{2}/g}$, which could produce seed black holes via the mechanism described in this paper.

We thank Shmulik Balberg, James Bullock, Renyue Chen, Phil Hopkins, Jun Koda, Sasha Muratov, Lisa Randall, Paul Shapiro, Stu Shapiro, Charles Steinhardt, and Naoki Yoshida for helpful discussions. We thank  especially Sasha Muratov for measuring concentration parameters at high redshifts in the FIRE runs and providing us with the resulting halo catalogs. This research is funded in part by DOE Grant \#DE-SC0011632, and by the Gordon and Betty Moore Foundation through Grant \#776 to the Caltech Moore Center for Theoretical Cosmology and Physics.

\bibliographystyle{utphys}
\bibliography{sidm_smbh_refs}

\end{document}